\documentclass[sigconf, nonacm]{acmart}

\newcommand\vldbdoi{XX.XX/XXX.XX}
\newcommand\vldbpages{XXX-XXX}
\newcommand\vldbvolume{14}
\newcommand\vldbissue{1}
\newcommand\vldbyear{2020}
\newcommand\vldbauthors{\authors}
\newcommand\vldbtitle{\shorttitle} 
\newcommand\vldbavailabilityurl{}
\newcommand\vldbpagestyle{plain} 

\usepackage{enumitem}

\DeclareMathOperator*{\argmin}{arg\,min}

\begin{document}

\title[MultiScope: Efficient Video Pre-processing for Exploratory Video Analytics]{MultiScope: Efficient Video Pre-processing for \\ Exploratory Video Analytics}
\author{Favyen Bastani}
\affiliation{%
  \institution{MIT CSAIL}
}
\email{favyen@csail.mit.edu}

\author{Sam Madden}
\affiliation{%
  \institution{MIT CSAIL}
}
\email{madden@csail.mit.edu}

\begin{abstract}
Performing analytics tasks over large-scale video datasets is increasingly common in a wide range of applications. These tasks generally involve object detection and tracking operations that require applying expensive machine learning models, and several systems have recently been proposed to optimize the execution of video queries to reduce their cost. However, prior work generally optimizes execution speed in only one dimension, focusing on one optimization technique while ignoring other potential avenues for accelerating execution, thereby delivering an unsatisfactory tradeoff between speed and accuracy. We propose MultiScope, a general-purpose video pre-processor for object detection and tracking that explores multiple avenues for optimizing video queries to extract tracks from video with a superior tradeoff between speed and accuracy over prior work. We compare MultiScope against three recent systems on seven diverse datasets, and find that it provides a 2.9x average speedup over the next best baseline at the same accuracy level.
\end{abstract}

\maketitle

\pagestyle{\vldbpagestyle}
\begingroup\small\noindent\raggedright\textbf{PVLDB Reference Format:}\\
\vldbauthors. \vldbtitle. PVLDB, \vldbvolume(\vldbissue): \vldbpages, \vldbyear.\\
\href{https://doi.org/\vldbdoi}{doi:\vldbdoi}
\endgroup
\begingroup
\renewcommand\thefootnote{}\footnote{\noindent
This work is licensed under the Creative Commons BY-NC-ND 4.0 International License. Visit \url{https://creativecommons.org/licenses/by-nc-nd/4.0/} to view a copy of this license. For any use beyond those covered by this license, obtain permission by emailing \href{mailto:info@vldb.org}{info@vldb.org}. Copyright is held by the owner/author(s). Publication rights licensed to the VLDB Endowment. \\
\raggedright Proceedings of the VLDB Endowment, Vol. \vldbvolume, No. \vldbissue\ %
ISSN 2150-8097. \\
\href{https://doi.org/\vldbdoi}{doi:\vldbdoi} \\
}\addtocounter{footnote}{-1}\endgroup
\ifdefempty{\vldbavailabilityurl}{}{
\vspace{.3cm}
\begingroup\small\noindent\raggedright\textbf{PVLDB Artifact Availability:}\\
The source code, data, and/or other artifacts have been made available at \url{\vldbavailabilityurl}.
\endgroup
}

\section{Introduction}

Over the last decade, improvements in machine learning methods, especially in convolutional neural networks (CNNs), have enabled numerous applications that involve querying large-scale video data. In particular, CNNs have been applied to accurately extract object detections (bounding box positions of objects) and tracks (sequences of bounding boxes over time) from video. Detections and tracks are used in virtually all video analytics tasks, such as in traffic planning to conduct turning movement counts (counting the number of cars turning in each direction in each time interval), in autonomous vehicle development to localize and track road signs, and in sports analytics to derive statistics from the motion of players and balls.

However, object detection methods are GPU-intensive: for example, on the \$10,000 NVIDIA Tesla V100 GPU, the YOLOv3 object detector~\cite{yolov3} can process $960 \times 540$ video frames at 100 frames per second (fps).
A user with a large volume of video that needs to be processed, say from hundreds of traffic cameras, would require one GPU for every 3-4 video feeds.
We could obtain some speedup by reducing the resolution and sampling rate at which the detector processes video, e.g., sampling only five $640 \times 480$ frames per second of video. However, the speedup that this provides is limited: accuracy drops off rapidly once the resolution is reduced far enough that the objects of interest occupy only a few pixels in the frame, or the sampling rate is low enough that objects move long distances between successively sampled frames.

Thus, recent work in the data management community has proposed systems that incorporate new optimizations for efficiently analyzing video~\cite{adascale,noscope,miris,chameleon}. However, these systems share several drawbacks, which we expand on in Section \ref{sec:background}: first, these systems generally each propose tuning a single parameter to provide a tradeoff between speed and accuracy, thereby optimizing performance only on a single dimension. Then, although the systems perform well against naive baselines that optimize on zero dimensions (always detecting objects at the full resolution and native video framerate), they provide a speed-accuracy curve that is oftentimes comparable to simply varying only the detector resolution and tracker sampling rate.
Second, several systems introduce query-driven optimizations that incorporate a slow query-specific execution phase
that introduces substantial per-query latency, thereby limiting practicality for exploratory analytics. For example, to optimize limit queries, BlazeIt~\cite{blazeit} first pre-processes video to build a query-agnostic index, but then conducts a search phase that needs to be repeated per-query and involves repeatedly applying an expensive object detector; the query-specific phase may require several minutes or even hours depending on the desired output cardinality and index precision, which is unacceptably long for exploratory queries.
Thus, a general-purpose video pre-processor that optimizes performance in multiple dimensions in order to efficiently extract all object tracks in a query-agnostic way would be ideal: it would allow users to efficiently answer exploratory queries by processing the extracted tracks without additional expensive ML inference.

To address this need, we developed MultiScope. We show that such a general-purpose pre-processor can in fact be achieved, while remaining competitive in speed even with query-driven approaches. MultiScope efficiently pre-processes raw video for downstream video analytics tasks by extracting all object tracks of a user-specified set of object categories from the video.
Users can then efficiently conduct exploratory analytics tasks by post-processing the tracks computed by MultiScope, without requiring further video decoding or ML inference; for example, a turning movement count query could be performed by counting the tracks that match each turning direction. To provide a superior speed-accuracy curve when inferring tracks in video, MultiScope (1) incorporates novel adaptations of two video analytics optimizations proposed in prior work; and (2) integrates these methods into a cohesive system that is able to simultaneously consider multiple parameters
to provide a greater speedup while introducing less error than prior approaches.

We now detail these two aspects of our method.
First, we develop novel adaptions of two optimizations, proxy models~\cite{noscope,blazeit} and reduced-rate tracking~\cite{miris}, to improve their robustness and speed by incorporating recent progress in computer vision techniques. Prior work in proxy models (NoScope~\cite{noscope}) trains fast classification models to input low-resolution video and estimate whether or not each frame contains at least one object detection; these models are then employed to skip execution of the slower detector model on frames where the proxy model has high confidence that there are no objects. However, many video datasets consist of busy scenes where there are objects in every frame, and proxy models provide no speedup. We extend the proxy model method to a multi-scale detection context~\cite{multiscale1}, where we use a proxy model to not only determine which frames contain objects, but also which spatial regions of frames contain objects. Then, even in videos of busy scenes, our method can still yield a speedup by only applying the slower detector in small windows of the frame that contain objects.

Prior work in reduced-rate tracking (Miris~\cite{miris}) proposes techniques to process video at substantially reduced sampling rates while still extracting accurate tracks. However, tracking models used in prior work are limited: they only consider matching detections between pairs of frames at a time (the previous frame and a new frame), and form tracks by creating chains of matches. Thus, the model cannot leverage useful cues such as object motion (e.g. velocity) that require analyzing multiple previous detections of an object. We instead employ a recurrent model that is able to account for information in multiple previous frames when matching detections in a new frame,
and address challenges to apply such a model in a reduced-rate tracking framework.

Second, we integrate these two novel techniques, along with a simple detector resolution optimization, into a cohesive system by applying a parameter tuning algorithm to choose multiple parameters across the three optimization methods, including the proxy model threshold, tracking sampling rate, and detector resolution.

We evaluate MultiScope on 7 diverse datasets on a task involving inferring all object tracks in video, and compare its performance in terms of speed and accuracy against three baselines: Chameleon~\cite{chameleon} (which derives a speedup primarily by tuning resolution and sampling rate), BlazeIt~\cite{blazeit}, and Miris~\cite{miris}.
Here, we apply BlazeIt and Miris under their respective query-agnostic execution modes.
We find that MultiScope consistently offers the best speed-accuracy tradeoff, and, when parameters are selected to obtain an accuracy within 5\% of the best-achieved accuracy (across the methods),
yields a 2.9x average speedup over the next fastest method. Moreover, we find that MultiScope outperforms BlazeIt even when BlazeIt is applied with its query-driven optimizations on limit queries with small output cardinalities (a type of query which BlazeIt specifically optimizes for) --- BlazeIt only outputs a limited number of query outputs, while MultiScope extracts all object tracks in the video, yet MultiScope achieves similar accuracy at a similar speed.

Our contributions are:
\begin{itemize}[noitemsep,topsep=0pt]
    \item We develop two novel video analytics optimization techniques, segmentation proxy models and recurrent reduced-rate tracking, that improve on prior work by incorporating new computer vision methods.
    \item We integrate these and other optimizations in a cohesive system that greedily tunes multiple parameters to select parameters providing high speed at every accuracy level.
    \item We show that MultiScope consistently offers the best speed-accuracy tradeoff against three baselines across seven diverse datasets when applied to infer all object tracks in video. Furthermore, we find that MultiScope is able to extract all tracks from video as fast as prior systems optimized for limit queries can process a single limit query.
\end{itemize}

\section{Background} \label{sec:background}

Recent systems proposed in the data management community for efficiently analyzing video have two critical drawbacks: they optimize performance only on a single dimension,
and they propose query-driven methods that introduce substantial per-query latency. In this section, we detail these drawbacks.

\smallskip
\noindent
\textbf{Multiple Avenues for Optimization.} The most substantial limitation in prior work is that the proposed systems focus on accelerating execution with one particular new method, and thereby ignore other important avenues for optimization. For example, NoScope~\cite{noscope} and BlazeIt~\cite{blazeit} apply an object detector at a fixed video resolution and sampling rate, and provide a tradeoff between speed and accuracy only through a threshold on the confidence of a proxy model that indicates the likelihood of there being at least one object in each frame of video. While proxy models are effective for processing videos of largely idle scenes that only rarely contain objects, they provide no benefit for extracting tracks from busy videos that have objects in every frame, since the proxy model cannot be used to skip any portion of such videos. Even for idle scenes, optimizing resolution and tracking sampling rate can provide an additional speedup with little effect on accuracy. Thus, considering multiple avenues for optimizing execution speed is crucial for a system to perform well across diverse video datasets.

While some prior work, such as Chameleon~\cite{chameleon}, propose video analytics systems that optimize performance over multiple parameters, these systems consider only video resolution and sampling rate, and do not address challenges in incorporating more recent optimizations such as proxy models and reduced-rate tracking.

In MultiScope, we propose a modular execution pipeline and parameter tuning architecture that consists of three components: an object detection module, a proxy model module, and a recurrent reduced-rate tracking module. The MultiScope parameter tuner selects parameters across these modules to provide the best tradeoff between speed and accuracy. By accounting for the multiple available avenues for optimizing execution, MultiScope substantially improves execution time over prior work.

\smallskip
\noindent
\textbf{Query-driven Methods.} Prior work such as BlazeIt and Miris incorporate a slow, query-specific phase that needs to be repeated for each query. In these systems, this phase involves applying an expensive object detector over selected frames, and thus introduces substantial per-query latency, limiting their practicality for exploratory analytics. If the overall cost of executing a few queries can nevertheless be reduced, this per-query latency may be worth it for some users. However, we will show that MultiScope can extract all object tracks from video in the same time that it takes BlazeIt to answer just one limit query.

\section{MultiScope}

MultiScope is a general-purpose video pre-processor for exploratory video analytics tasks that involve object detections or tracks. Given a video dataset, MultiScope efficiently and accurately extracts all object tracks from the video: its output is a set of tracks $\{ t_1, \ldots, t_n \}$, where each track $t_i = (C_k, \langle d_1, \ldots, d_{m_i} \rangle)$ is a unique object of some category $C_k$ (e.g., car or pedestrian) visible in the video represented as a sequence of detections, and each detection $d_i = (t, x, y, w, h)$ specifies a timestamp $t$ and a bounding box where the object appears. MultiScope is thus similar to work in multi-object tracking~\cite{famnet,gsm,cttrack}, but provides substantially faster execution speed by incorporating several diverse optimizations. After pre-processing video with MultiScope, users can rapidly answer queries by post-processing the computed tracks, without needing additional video decoding or ML inference. Four example queries that can directly be answered in traffic camera video from extracted object tracks are: (1) find cars that decelerate at $5 \text{m}/\text{s}^2$ or more (hard braking); (2) find frames with at least three buses and three cars; (3) find the average number of cars visible in the video over time; (4) find the average number of \emph{unique} cars over time (i.e., the traffic volume).

\subsection{Workflow}

\begin{figure}
\begin{center}
	\includegraphics[width=0.8\linewidth]{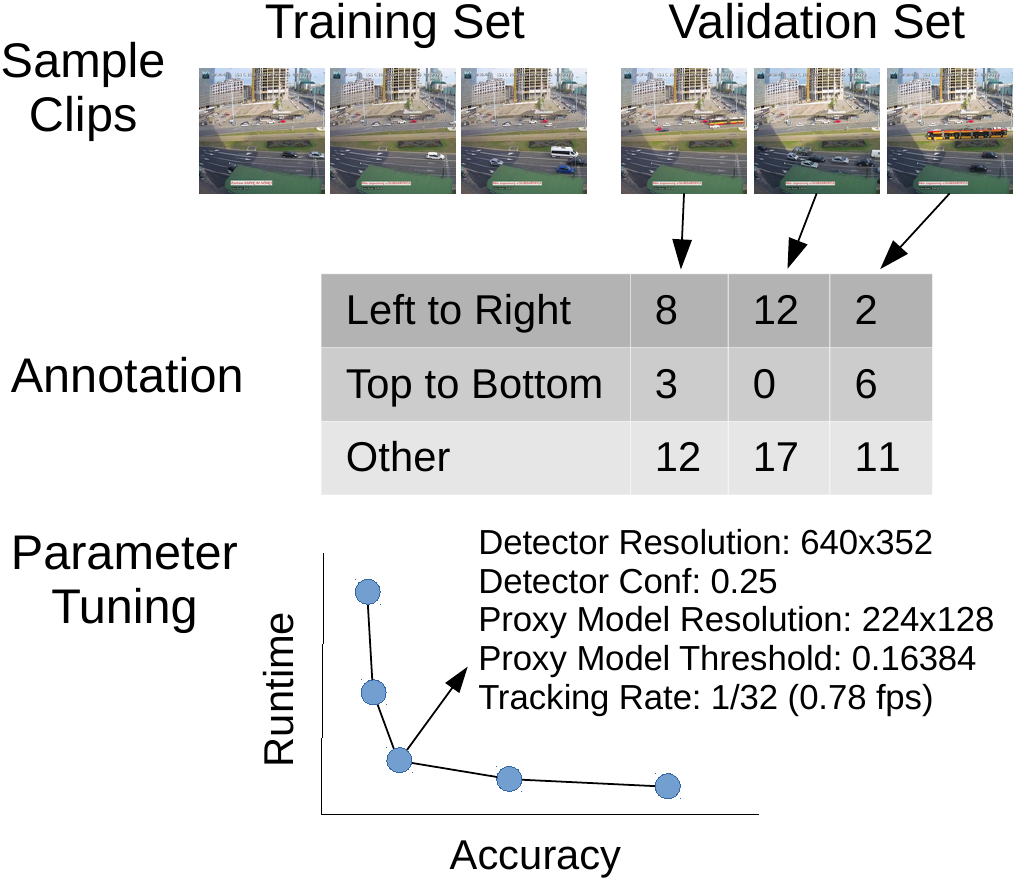}
\end{center}
	\caption{Overview of the MultiScope workflow. 60 one-minute clips are sampled from the dataset to form training and validation sets. The user provides ground truth data in each validation clip, e.g. by annotating counts of objects following each of three spatial patterns. The tuner outputs a sequence of parameter configurations that offer a tradeoff between speed and accuracy. The user selects one configuration (one point along curve) to apply on the entire dataset.}
\label{fig:overview}
\end{figure}

Before detailing MultiScope's design, in this section we first describe the workflow of applying MultiScope on a new video dataset (Figure \ref{fig:overview}). Users first sample training and validation sets from the dataset, which each consist of many sampled clips of a certain length --- in our implementation, we sample one hour of video consisting of 60 one-minute clips for each of the training and validation sets. MultiScope uses the training set to train proxy models, and uses the validation set to select parameters.

The user then provides a metric and corresponding ground truth for evaluating the accuracy of tracks extracted by various MultiScope parameter configurations in each validation clip.
This ground truth could correspond to the outputs of an automatic ``oracle'' pipeline where we apply an object detector and tracker at the native video resolution and framerate. This strategy is proposed in NoScope~\cite{noscope} to avoid the need for hand-labeling. However, we will show in Section \ref{sec:eval} that without a metric that leverages hand-labeled data, the degree of error in tracks produced automatically by these noisy oracles is unbounded, making it impossible for the user to judge what accuracy level is acceptable. Thus, alternatively, the user could hand-label the ground truth. For example, the user can provide track labels, where the user draws bounding boxes around all objects of interest in each clip, and labels sequences of boxes that correspond to the same objects.
However, this type of data is tedious to hand-label. As a third alternative, the user can provide labels for the number of unique objects that are visible in the clip, broken down based on certain spatial patterns (e.g. count the number of cars that pass a junction under each turning direction). These labels can be annotated more quickly --- on some datasets, just 20 minutes to label a one-hour validation set --- and we will show that they still enable MultiScope to produce accurate tracks.

The MultiScope parameter tuner will then experiment with various parameter configurations and evaluate the speed and accuracy when executing the pipeline under each configuration over the validation set. The tuner begins with the slowest possible configuration (which may or may not yield the highest accuracy),
and then greedily selects parameters that yield speedups with the smallest reductions in accuracy. The output of this process is a speed-accuracy curve, where each point along the curve corresponds to one parameter configuration. The user can then select a point on the speed-accuracy curve, and MultiScope will extract tracks over the entire video dataset using the corresponding parameters.

\subsection{Architecture}

\begin{figure}
\begin{center}
	\includegraphics[width=0.8\linewidth]{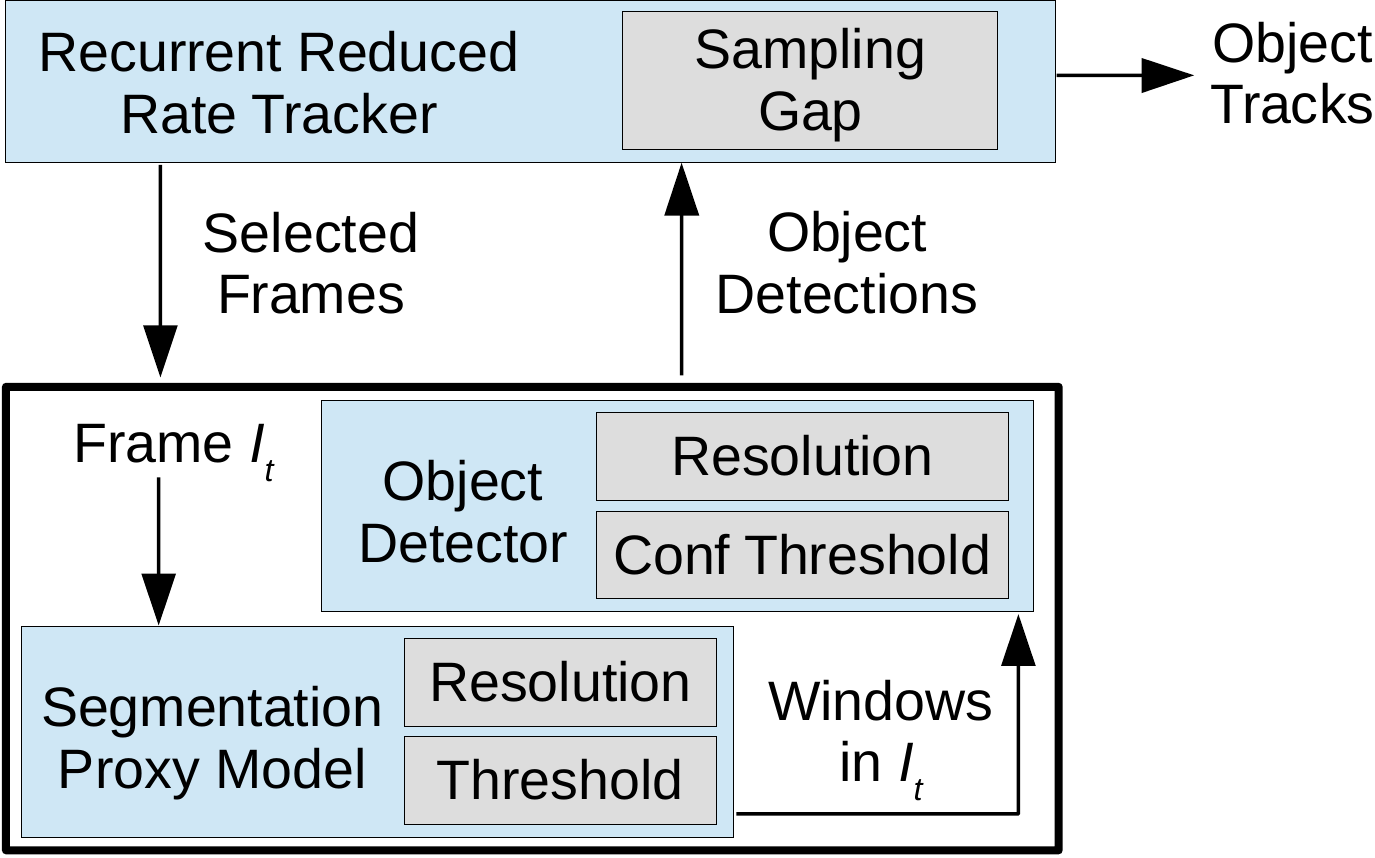}
\end{center}
	\caption{MultiScope execution pipeline architecture. The tracker selects which frames to process. To detect objects in each sampled frame, the segmentation proxy model determines which windows of the frame may contain objects, and the detector runs in those windows. Parameters selected by the tuner are shown in grey.}
\label{fig:arch}
\end{figure}

We now introduce the MultiScope architecture (Figure \ref{fig:arch}) at a high level. The execution pipeline consists of three modules, where each module exposes several parameters that influence speed and accuracy.
First, a segmentation proxy model determines which frames and which parts of frames contain objects, so that a more expensive object detector can be executed only on those regions. This module is configured with the input resolution of the proxy model, and a threshold on the proxy model confidence that determines how confident the model must be before skipping the processing of portions of frames. Second, the detection module applies an object detection model, and is configured with the model architecture (e.g., YOLO~\cite{yolo} or Mask R-CNN~\cite{maskrcnn}), input resolution, and detection confidence threshold. Lastly, sitting on top of the other two modules, the recurrent reduced-rate tracking method decides which frames should be processed for computing object detections, and groups detections of the same object across different frames to produce object tracks. The tracking module is configured with a sampling gap that specifies the rate at which frames should be processed.

Additionally, MultiScope includes a parameter tuner that outputs a speed-accuracy curve of parameter configurations using a greedy algorithm. After training and validation sets are sampled, and the evaluation metric for the validation clips is provided by the user, MultiScope initializes by training a range of proxy models and a recurrent tracking model, and then executing the tuner.

Below, we introduce our novel segmentation proxy model method in Section \ref{sec:msproxy} and our recurrent reduced-rate tracking method in Section \ref{sec:mstrack}. We then detail the MultiScope tuner in Section \ref{sec:msoptimize}.

\subsection{Segmentation Proxy Model} \label{sec:msproxy}

\begin{figure}
\begin{center}
	\includegraphics[width=\linewidth]{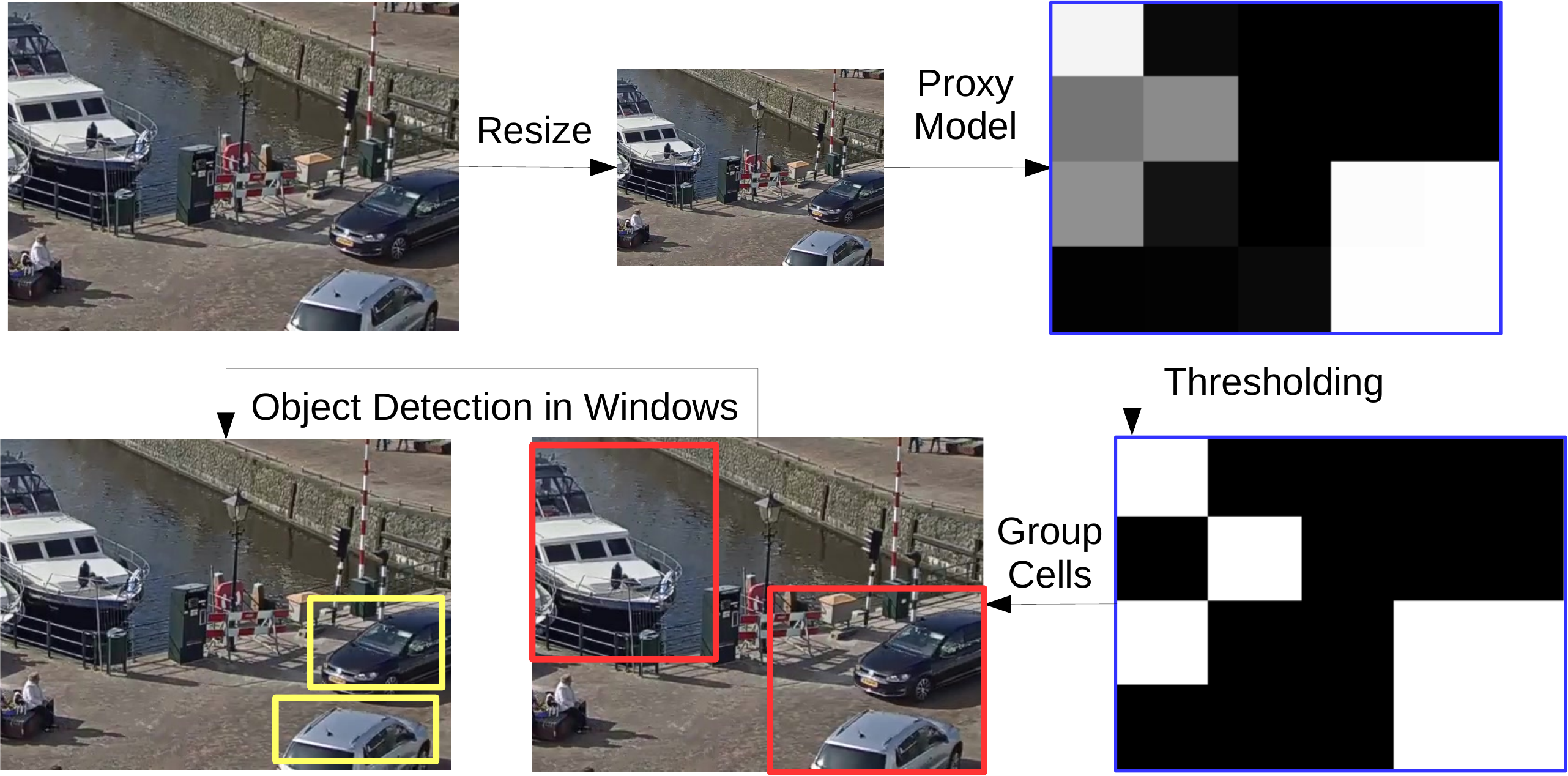}
\end{center}
	\caption{Summary of our novel segmentation proxy model method. A proxy model inputs a video frame at a low resolution, and scores each cell in the frame with the likelihood that the cells intersects a detection. Positive cells after thresholding are grouped into rectangular windows, and the object detector is applied only in those windows.}
\label{fig:segment}
\end{figure}

In prior work, proxy models are applied to determine which frames should be processed to compute object detections. For example, NoScope~\cite{noscope} trains a proxy model to classify whether or not a video frame contains at least one object. Then, NoScope skips object detection processing on frames where the proxy model has sufficiently low confidence. The proxy model is substantially faster than the object detector since it inputs video at a lower resolution, and, to a lesser extent, since it employs a shallower model architecture; thus, this yields a speedup in videos where a large fraction of frames contain zero objects.
However, we find that many video datasets contain relevant objects in every frame --- for example, video of a traffic junction may continuously contain cars if the junction is busy. Classification proxy models provide no speedup in such videos since no frames can be skipped entirely.

Intuitively, though, proxy models still could provide a benefit by identifying {\it regions} of frames that contain no objects, and skipping object detection processing on those regions. Then, if the video contains many segments where the camera frame is sparsely (spatially) populated by objects, this method can provide a substantial speedup: as long as the proxy model can accurately distinguish regions with objects from regions without objects at a lower resolution than that at which the object detector can accurately compute bounding box detections, then although the detector must still be applied on each frame, it can be applied only in small windows where the proxy model determines that objects are present.

In this section, we detail our novel segmentation proxy model method that implements this idea. We employ a segmentation CNN model architecture, which processes an image and outputs a score at each grid cell of pixels in the image (e.g., every $32 \times 32$ cell). We train the model to classify whether each cell intersects at least one detection. Then, during inference, we aim to only apply the detector on ``positive'' cells where the proxy model has high confidence, i.e., where the score exceeds a threshold parameter $B_\text{proxy}$.

However, there are several challenges with applying the detector in this way. First, objects may span multiple adjacent cells, and the object detector can only be efficiently applied on GPUs on rectangular inputs. Thus, after using the proxy model to determine a set of positive cells that may contain detections, we must aggregate these cells together into rectangular groups such that the rectangles cover all of the positive cells; we can then apply the object detector in these rectangles. Second, the object detector is much slower when applied on variable-dimension inputs, since high detection performance on GPUs relies heavily on batching many equal-dimension inputs together. While input padding is often used to ensure all inputs are the same dimension, in our scenario, this would erase the time savings from applying the detector on small regions of a video frame. Instead, we develop an algorithm that determines ahead of time a small number (three in our implementation) of fixed window sizes, and initializes the detector on the GPU to execute at each of those sizes. Then, during inference, for each frame of video where we need to compute detections, we select rectangles sized at one of the pre-selected window sizes to cover the positive cells determined by the proxy model.
Figure \ref{fig:segment} summarizes our approach.

\smallskip
\noindent
\textbf{Model Architecture.} We employ a simple, standard segmentation CNN architecture for the proxy model. Our model consists of a five-layer encoder followed by a two-layer decoder. The encoder inputs the video frame, and applies a series of five strided convolutional layers, producing features at 1/32 the resolution of the input. The decoder applies two additional convolutional layers, and its output is a classification score at each $32 \times 32$ cell of the input image indicating the likelihood that the cell intersects an object. We opt for a $32 \times 32$ cell size since objects are usually comparable or larger in size, and since this yields few enough cells so that the data does not become unwieldy when we group cells into rectangular windows.

\smallskip
\noindent
\textbf{Training.} As in prior work, we use the object detection outputs of a {\it best-accuracy} parameter configuration $\theta_\text{best}$ as rough labels for training the segmentation proxy model, where the model should output a score close to 1 at cells that intersect a detection, and 0 at other cells.
$\theta_\text{best}$ is a configuration of parameters in the MultiScope pipeline that provides the best accuracy (which may still be far from 100\%).
We detail the selection of $\theta_\text{best}$ in the next sub-section.

Before training the proxy model, we first compute object detections $D^{(t)} = \{ d^{(t)}_1, \ldots, d^{(t)}_n \}$ using $\theta_\text{best}$ on each frame $I_t$ in the training set of video. Then, during training, we generate input-output training examples by first sampling a frame $I_t$ from the training set where $|D^{(t)}| > 0$, i.e., at least one detection was output by the best-accuracy configuration. We construct classification labels for $I_t$ corresponding to the detections in $D^{(t)}$: at each $32 \times 32$ cell, the label is 1 if there is some detection $d^{(t)}_i \in D^{(t)}$ that intersects the cell, and 0 otherwise.

Prior to training the proxy model, we cannot be certain how accurate the model will be at a certain model input resolution --- different resolutions, such as inputting $416 \times 256$ or $224 \times 128$ frames (which yield $13 \times 8$ and $7 \times 4$ output grids, respectively), may provide tradeoffs between speed and accuracy.
Thus, we train several proxy models at various pre-determined resolutions (a range of five resolutions in our implementation), and leave the resolution as a parameter for the tuner to select from this range. Although we train multiple models, the training phase requires <10 minutes before convergence of all five models because all input resolutions are much lower than the native video resolution.

\smallskip
\noindent
\textbf{Best-accuracy Configuration Selection.} As discussed above, MultiScope requires the outputs of a best-accuracy configuration $\theta_\text{best}$ to train proxy models. 
$\theta_\text{best}$ is simply a selection of parameters for the MultiScope pipeline that yields highest accuracy over the validation set on the user-provided metric, which we can use to obtain object detections and tracks in the training set of video for the purpose of training models. These detections and tracks may contain errors, but nevertheless correspond to the best accuracy we can obtain through automatic labeling. To select these parameters, we begin by evaluating the accuracy of the slowest possible configuration on the validation set (using the metric and ground truth provided by the user), i.e., the configuration with no segmentation proxy model, maximum object detector resolution, and maximum sampling rate. Then, we repeatedly reduce the detector resolution by 15\% on each dimension and re-evaluate accuracy until the accuracy decreases, and keep the resolution providing the best achieved accuracy. We then reduce the sampling rate in a similar procedure: we repeatedly reduce the rate by 2x and re-evaluate accuracy until accuracy drops. This procedure is crucial because we find that accuracy is oftentimes higher at lower resolutions.
We do not consider employing a proxy model for $\theta_\text{best}$ since the proxy model never improves accuracy, and because at this stage the proxy models have not yet been trained. Since the MultiScope tracking model has also not yet been trained, we use the heuristic SORT tracker~\cite{sort} in $\theta_\text{best}$, which tracks objects based on bounding box overlap.

\smallskip
\noindent
\textbf{Inference.} The proxy model inference procedure is configured by two parameters: the model input resolution (which determines which of the several trained proxy models to use) and a confidence threshold $B_\text{proxy}$ on the segmentation outputs. We will discuss selecting these parameters in Section \ref{sec:msoptimize}. During inference, on each frame of video, we apply the proxy model to compute classification scores on each $32 \times 32$ cell. After thresholding the scores by $B_\text{proxy}$, we derive a binary grid consisting of a (possibly empty) set of positive cells where the output scores exceeded $B_\text{proxy}$. These cells are ones in which we must apply the object detector. The final step in the inference procedure is to select rectangular windows of the video frame in which to apply the detector. These windows should cover the positive cells, but should do so tightly to minimize the execution time needed to apply the detector over the windows. We will discuss this final step in two sections below: grouping cells into rectangular windows of certain sizes during inference, and determining ahead of time the fixed set of window sizes at which we will run the object detector.

\smallskip
\noindent
\textbf{Grouping Cells during Execution.} On a frame $I_t$, the proxy model yields a set of positive cells $X^{(t)} = \{ x^{(t)}_1, \ldots, x^{(t)}_n \}$. We assume we are given a fixed set of window sizes at which the detector will run, $S = \{ (w_1, h_1), \ldots, (w_k, h_k) \}$, as well as the detector execution time of each size, $T_{w,h}$. We will detail how we decide on $S$ prior to execution in the next section. Our aim is to find a set of rectangular windows $R_t = \{ r_1, \ldots, r_m \}$ that covers all of the cells in $X^{(t)}$. Here, each $r_i$ is a 4-tuple $(r_i.x, r_i.y, r_i.w, r_i.h)$ that specifies the position and size of the rectangle, where $(r_i.w, r_i.h) \in S$. Let $est(R_t) = \sum_{r_i \in R_t} T_{r_i.w, r_i.h}$ be the estimated execution time of applying the detector in these windows. Then, in particular, we want the optimal set of rectangles $R^{*}(I_t; S)$ that minimizes $est(R_t)$.

We approximate the optimal solution using a density-based agglomerative clustering method. Essentially, each cluster corresponds to a proposed rectangle, and our clustering procedure will repeatedly greedily check whether two clusters can be combined to form a merged cluster that would be faster to process through the object detector than processing the two original clusters separately. We initialize a cluster $C_i$ for each connected component of positive cells. Note that we do not create one cluster per cell: because objects may span across cells, we want connected components of positive cells to be contained inside the same rectangle. We then iterate over the clusters. For each cluster $C_i$, we identify its closest neighbor $C_j$. We create a proposed merged cluster $C_\text{merged}$, and identify the smallest-area window size $(w, h)$ that contains the bounding box of $C_\text{merged}$  (i.e., can cover all of the positive cells in $C_\text{merged}$). For every other cluster $C_k$, we check if we can add $C_k$ to $C_\text{merged}$ without needing a larger window size. Finally, to determine whether to incorporate the proposed cluster $C_\text{merged}$, we compare the estimated time to process $C_\text{merged}$ (i.e., $T_{w,h}$) with the sum of the time to process the individual clusters (based on the smallest-area window sizes in $S$ that contain the bounding box of each cluster). If the execution time for $C_\text{merged}$ is smaller, then we remove the individual clusters and add $C_\text{merged}$. We repeatedly loop over the clusters until we perform a pass without any new merges.

After clustering terminates, we construct a set of rectangular windows $R$ by creating one rectangle for each cluster. 

\smallskip
\noindent
\textbf{Determining Fixed Set of Window Sizes.} Prior to execution, we decide the fixed set of window sizes $S = \{(w_1, h_1), \ldots, (w_k, h_k)\}$ (of a predetermined cardinality $k = 3$ that is configured based on available GPU memory) at which to run the object detector, so that we can take advantage of the substantial speed savings from batch execution of the detector on the GPU. The optimal set of window sizes $S^{*}$ is the set that yields the highest execution speed in expectation over video segments sampled uniformly from the dataset. When computing $S$, to simplify the problem, we assume that our proxy model performs perfectly (i.e., positive cells correspond directly to the locations of object detections).
Then, it follows that $S^{*} = \argmin_S \sum_{t} est(R^{*}(I_t; S))$. Intuitively, $S^{*}$ should tightly cover the locations of objects in video frames, so that on most frames, we can apply the detector only in small windows where objects are present. Our aim is to pick $S$ so that the resulting execution speed is close to that of $S^{*}$.

To do so, we first initialize $S$ to only contain the size corresponding to the entire video frame --- we always include this maximum window size in the set so that the option of simply applying the detector on the entire frame is always available. We assume for now that we have a function $tot\_time(S)$ that estimates the expected execution time of a set $S$; we will detail our approach for implementing $tot\_time(S)$ below. Then, to pick the remaining window sizes, we employ a greedy approach that iteratively selects one window size at a time to add to $S$: on each iteration, we select the size $(w, h)$ that minimizes $tot\_time(S + \{(w, h)\})$. We try all possible dimension $(w, h)$ that are smaller than the video frame and where $w$ and $h$ are both multiples of 32.

It remains to implement $tot\_time(S)$. For each image $I_t$ in the training set of video, we will compute an estimate $im\_time(S, I_t)$ of the execution time for that frame, and then compute $tot\_time(S)$ as the sum $\sum_t im\_time(S, I_t)$ over the training set. To simplify $im\_time(S, I_t)$, we assume that the proxy model will perfectly produce positive cells matching the positions of object detections computed by $\theta_\text{best}$ in $I_t$. Then, $im\_time(S, I_t)$ should equal $est(R(I_t; S))$, where we define $R(I_t; S)$ as the set of rectangular windows computed using our method for grouping cells detailed above.

\subsection{Recurrent Reduced-Rate Tracking} \label{sec:mstrack}

Prior work (Miris~\cite{miris}) applies reduced-rate tracking to speed up the grouping of frames into object tracks in a query-driven context where it is assumed that a query consisting of a predicate on object tracks is provided, e.g., finding tracks of cars in traffic camera footage that travel north to south through a junction. Reduced-rate tracking applies an object detector at greatly reduced sampling rates (e.g. 1 fps instead of 25 fps), and attempts to accurately recover groups of detections of the same object.

The method in Miris has two limitations. First, it relies on capturing additional detections at high sampling rates to predict the position and time where a track first becomes visible, along with that where it is no longer visible. This step is needed since, when tracking at reduced rates, the first and last detections observed in the track may be a second before or after the object actually enters or leaves the camera frame; thus, for example, a car that traveled north to south through a junction may first be seen in the middle of the junction, resulting in the track being excluded from the north-to-south query. While the procedure of ``refining'' tracks by computing additional detections is effective for extracting tracks under predicates that only select small subsets of tracks, it becomes cost-prohibitive when extracting all tracks from video, which is the objective of MultiScope. Second, Miris employs a graph neural network (GNN) model that only compares detections in two consecutive frames at a time to produce probabilities that each pair of detections across the frames correspond to the same object. Although this simplifies the training process and the overall tracking architecture, it limits accuracy when tracking at low sampling rates, since, unlike a recurrent model, the GNN model cannot exploit motion and other patterns observed in multiple previous frames when matching detections in a new frame to existing track prefixes.

We propose a recurrent reduced-rate tracking method in MultiScope to address these challenges. First, we train a recurrent tracking network similar in design to architectures used in state-of-the-art multi-object tracking methods in the computer vision literature. A recurrent model inputs a sequence of data step-by-step, and produces a corresponding output sequence; here, the recurrent tracker processes video one frame at a time, inputting the new object detections in each successive frame and outputting the associations between those detections and previously observed tracks. Unlike prior work, though, we use a novel specialized training process that we propose to ensure the model is robust when applied at varying reduced sampling rates. Second, rather than apply the track refinement methods introduced in Miris to accurately localize the start and end of each track, we propose estimating the start and end position based on the average path of the most similar tracks computed in the training set. To do so efficiently, we cluster and index the training set tracks ahead of time so that similar tracks can be looked up quickly during inference.

\smallskip
\noindent
\textbf{Model Architecture.} We use a recurrent tracking model architecture that is comparable to that used in recent work in multi-object tracking~\cite{rnnmot2}. The tracker inputs a video sequence $\langle I_1, \ldots, I_n \rangle$ and object detections computed by the detector over the sequence, where $D^{(t)} = \{ d^{(t)}_1, \ldots, d^{(t)}_{n_t} \}$ denotes the detections computed in frame $I_t$. The goal of the tracker is to assign a track ID to each detection so that detections of the same object are all assigned the same track ID. To apply the model, we iterate over the frame sequence, maintaining a set of active track prefixes $A = \{ s_1, \ldots, s_m \}$ computed up until the previous frame. Processing begins on $I_1$, where we initialize a new track prefix $s_i = \langle d^{(1)}_i \rangle$ for each detection in $D^{(1)}$. On a successive frame $I_t$, we apply a neural network model to compute a score $p^{(t)}_{i,j}$ indicating the likelihood that each detection $d^{(t)}_j \in D^{(t)}$ corresponds to each track prefix $s_i$. We apply the Hungarian algorithm  to match detections with tracks based on these scores, and add each detection to the track that it matches with. If a detection $d^{(t)}_j$ does not match with any track, we initialize a new track prefix $s = \langle d^{(t)}_j \rangle$ and add it to $A$.

Our model has three components. First, we compute detection-level features that describe each detection $d = (x, y, w, h)$. This component applies a CNN to process the image content of the detection (i.e., the portion of the frame contained in the bounding box that $d$ specifies) into a compact feature vector representation. Then, the detection-level feature vector for $d$ is the concatenation of the representation computed by the CNN with the 4D spatial bounding box coordinates $(x, y, w, h)$. Second, we compute track-level features that describe each track prefix. A track $s = \langle d_1, \ldots, d_n \rangle$ is a sequence of detections, so it is natural to use an RNN for this component. We compute the detection-level features for each detection in $s$, and apply an RNN over those features. We use the final output of the RNN as the track-level features. Third, we compute the scores $p^{(t)}_{i,j}$ matching the detection $d^{(t)}_j$ with $s_i$. This component applies a matching network consisting of several fully connected layers that inputs the detection-level features of $d^{(t)}_j$ and the track-level features of $s_i$, and outputs a single score.

\smallskip
\noindent
\textbf{Training.} Typically, tracking models are trained on ground truth labels annotated at the full video framerate. However, we do not have ground truth labels, and we require our tracker to be effective at several possible reduced sampling rates. To address the first issue, as with training the segmentation proxy models, we use outputs of the best-accuracy configuration $\theta_\text{best}$ from Section \ref{sec:msproxy} as a rough ground truth for training. Let $S^{*} = \{s^{*}_1, \ldots, s^{*}_n\}$ be the set of tracks computed by $\theta_\text{best}$ over the training set of video.

To train a model that will be robust even when track prefixes are captured at various reduced sampling rates, we construct training examples that consist of detection sequences sampled from some $s^{*}_i \in S^{*}$ with diverse spacing. During inference, we will process video at some sampling gap, where each gap is a power of two, and a gap $g$ specifies that one out of every $g$ frames should be processed ($g = 1$ corresponds to processing the video at its native framerate). Let $G = \langle 1, 2, 4, \ldots, 2^n \rangle$ be the maximal gap sequence, i.e., we will not track at sampling gaps higher than $2^n$ during inference. Then, to create a training example, in addition to sampling a track $s \sim S^{*}$, we sample a gap $g \sim G$. We then iteratively sub-sample detections from $s$, beginning from its first detection, such that each following detection is at least $g$ frames after the previous detection.

We have shown how to generate examples for training the model. However, if we do not provide any temporal information to the model, then when matching a detection $d$ with a track prefix $s$, the model cannot, for example, predict the position of $s$ at the current frame based on its velocity in preceding frames, since doing so would necessitate multiplying this velocity by the time elapsed between $d$ and the last detection in $s$. Thus, we augment the detection-level features of $d$ to include the number of frames $t_\text{elapsed}$ between $d$ and some preceding detection, to enable the model to account for temporal information. When computing track-level features of $s = \langle d_1, \ldots, d_n \rangle$, the value of $t_\text{elapsed}$ for $d_i$ is the number of frames between $d_{i-1}$ and $d_i$. When computing detection-level features for a detection $d^{(t)}_j$ in the current frame $I_t$, $t_\text{elapsed}$ is the number of frames between $I_t$ and the previously processed frame. Providing this additional input enables the model to more robustly compute scores $p^{(t)}_{i,j}$ when using reduced-rate tracking.

\smallskip
\noindent
\textbf{Inference.} Before applying the tracker on a video dataset, we assume that the MultiScope tuner (Section \ref{sec:msoptimize}) has selected a gap $g \in G$.
We decode video at a framerate corresponding to the gap $g$ to obtain a sequence of frames $\langle I_0, I_g, I_{2g}, \ldots \rangle$. On each frame $I_t$, we first compute a set of object detections $D^{(t)}$ in the frame using the segmentation proxy model and object detector components. We then apply the tracker model to compute scores $p^{(t)}_{i,j}$ between detections $d^{(t)}_j \in D^{(t)}$ and track prefixes $s_i$, apply the Hungarian algorithm to derive a matching between detections and tracks based on these scores, and update the tracks based on the matching.

Note that we do not use the variable rate selection technique in Miris, which tracks objects not at a single gap $g$ but at a variable gap that changes depending on the tracker confidence. We conducted preliminary experiments to evaluate the effectiveness of the variable technique when incorporated into MultiScope, but found the accuracy comparable to simply using a fixed gap when employing our recurrent tracking model.

\begin{figure}
\begin{center}
	\includegraphics[width=\linewidth]{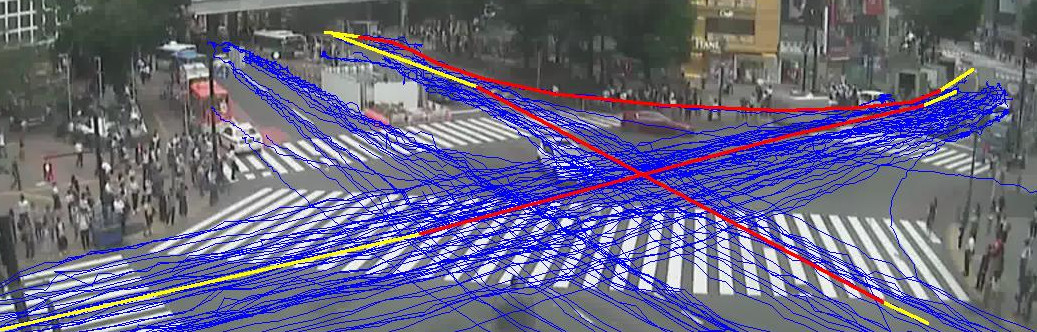}
\end{center}
	\caption{Rather than refine the start and end of tracks captured at high sampling gaps by processing additional frames, we estimate the start and end position based on similar tracks seen in the training set. Here, blue lines show clusters computed from the training set tracks, red lines show tracks captured at a high sampling gap, and yellow lines show extensions to the tracks added by our refinement method.}
\label{fig:refine}
\end{figure}

\smallskip
\noindent
\textbf{Refinement.} The recurrent reduced-rate tracking method that we have introduced improves accuracy when tracking objects at reduced sampling rates. However, one crucial issue with tracks extracted at low sampling rates is that their first and last detection will be offset from the start and end of the object's actual trajectory. This is most problematic when extracting tracks in video captured from fixed cameras, because oftentimes, video analytics tasks over such video involve spatial predicates on the first and last detections in tracks: for example, performing a turning movement count in video of a traffic junction necessitates counting the number of tracks of cars that start and end at each pair of roads incident on the junction. Simply extrapolating a track to the frame boundary is insufficient since tracks may start or end at the horizon, or anywhere in the frame due to an occlusion.

Prior work~\cite{miris} proposes processing additional frames to refine tracks. However, this becomes cost-prohibitive when extracting all tracks from video.
Instead, we observe that in fixed video (where the offset is most problematic), we can generally estimate the start and end of a track by comparing the portion of a track captured at a low sampling rate against known tracks captured at the native framerate. We show an example in Figure \ref{fig:refine}.
Then, a simple algorithm for refining tracks would be to find the k-nearest neighbors $knn(s)$ of a track $s$ among tracks captured by the best-accuracy configuration, $S^{*}$, and estimate the start and end of $s$ as the median start and end of tracks in $knn(s)$. 

However, computing pairwise distances between tracks grows more expensive as $S^{*}$ increases in size, and while we could sample a subset of tracks from $S^{*}$ to use for k-nearest neighbor computation during inference, this reduces accuracy for infrequently-occurring paths. To speed up k-nearest neighbor lookups, we cluster tracks in $S^{*}$ prior to inference, and construct a spatial index over cluster centers (which are paths). Clustering reduces the number of distances that must be computed, so that redundant tracks that follow the same path in the video are merged into one cluster.

In our approach, we begin by clustering the tracks in $S^{*}$ using DBSCAN. To apply DBSCAN, we must define the distance metric between tracks, as well as how to compute the center of a cluster. We use a simple distance metric: given $s_1$ and $s_2$, we compute $N$ points evenly spaced along each track, yielding sequences of points $P(s_1)$ and $P(s_2)$, and define $d(s_1, s_2) = \frac{1}{N} \sum_{i=1}^N \text{eucl}(P(s_1)[i], P(s_2)[i])$, i.e., as the average distance between corresponding points. Here, $\text{eucl}(p_1, p_2)$ is the Euclidean distance between 2D points. In our implementation, $N = 20$. We define the center of a cluster of tracks $C = \{ s_1, \ldots, s_n \}$ as a path $s = \langle p_1, \ldots, p_N \rangle$ of $N$ points, where $p_i$ is the average of points in $\{ P(s)[i] | s \in C \}$, i.e., the average position of the $i$th evenly spaced point computed among tracks in the cluster.

DBSCAN produces a set of clusters. We then build a simple spatial grid index over the cluster centers, where each grid cell maps to a list of paths that pass through that cell. During inference, we approximate the k-nearest neighbor lookup for a computed track $s = \langle d_1, \ldots, d_n \rangle$ as follows. We first use the index to identify several cluster centers that pass close to $d_1$ and $d_n$. We compute the distance between each of those cluster centers and $s$, and keep the $k = 10$ closest cluster centers, where a cluster of $n$ tracks counts $n$ times.
Finally, we extend $s$ with a start and end detection computed by taking the median coordinate on each dimension of the start and end points across the clusters, weighted by cluster sizes.

\subsection{Joint Parameter Tuning} \label{sec:msoptimize}

Each module in the MultiScope pipeline is governed by several parameters. The {\it MultiScope tuner} is responsible for selecting values for each of these parameters. Specific parameters include:

\begin{itemize}
    \item The detection module is configured by the object detector model architecture (e.g. YOLOv3 or Mask R-CNN), input resolution, and confidence threshold.
    \item  The proxy model module is configured by the input resolution and confidence threshold. 
    \item The tracking module is configured by the sampling gap $g$. 
\end{itemize} 

The output of the tuner is a sequence of configurations $\Theta = \langle \theta_1, \ldots, \theta_n \rangle$, where each $\theta_i$ represents a setting of the above parameters.
$\Theta$ forms a speed-accuracy tradeoff curve, and the goal of the tuner is for it to be as close as possible to the Pareto-optimal frontier of the space of all possible configurations. After the tuning process completes, the user will pick a configuration $\theta_i$ along the output curve to use for execution over the rest of the dataset.

A brute force approach would be to compute the validation set accuracy and speed of every possible configuration, under some search granularity for real-valued parameters. However, depending on the search granularity, there may be tens of thousands of potential configurations, since the search space is exponential in the number of parameters.

Instead, we adopt an iterative greedy algorithm. Our method is similar to that used in prior work such as in Chameleon~\cite{chameleon}, but incorporates a modular architecture to accommodate the increased number of parameters in our execution pipeline. We begin with a slow but accurate configuration, and on each iteration, we independently update different subsets of parameters to offer a speedup, test the accuracy of each update, and choose the update that offers the best accuracy.

We initialize $\theta_1 = \theta_\text{best}$ as the best-accuracy configuration (Section \ref{sec:msproxy}), i.e., the parameters that yield the best possible accuracy on the validation set, but poor execution speed. On each iteration, given the current configuration $\theta_i$, the tuner queries each module and requests a new configuration where the parameters for that module have been updated to provide an overall speedup of approximately $S$ over $\theta_i$ (we use $S=30\%$ in our implementation, meaning the next configuration should be around 30\% faster). This yields three candidate configurations $\theta_{i+1,\text{detection}}, \theta_{i+1,\text{proxy}}, \theta_{i+1,\text{tracking}}$, where each reflects updates to the parameters of one module that provide the desired overall speedup. Then, the tuner executes each candidate configuration over the validation set to measure accuracy. We set $\theta_{i+1}$ to the best-accuracy candidate, and iteratively repeat this procedure. By repeatedly choosing parameter changes that provide the smallest drop in accuracy for a fixed desired speedup, this approach produces an output $\Theta$ that approximates the Pareto frontier. If there are $m$ modules and we seek a curve of $n$ configurations, then the algorithm uses $O(mn)$ trials on the validation set ($n$ iterations where we test $m$ candidates on each iteration).

The MultiScope tuner executes in two phases. First, during a caching phase, each module caches information that it will need to answer the tuner requests for next candidate configurations (i.e., to compute a configuration $\theta_{i+1,\text{module}}$ that is $S$ faster than $\theta_i$).  Then, during the tuning phase, we apply the greedy algorithm detailed above that produces $\Theta$.
Below, we detail how each module operates during the caching and tuning phases.

\subsubsection{Detection Module}

Because the object detector is generally the slowest component of the pipeline, increasing object detection speed by $S$ usually corresponds to an overall speedup of almost $S$. Then, at a high level, our strategy in the detection module during the tuning phase is to simply find the object detector configuration that offers the highest accuracy among configurations that are at least $S$ faster than the previous configuration.

Let $A = \{ a_1, \ldots, a_n\}$ be the set of model architectures (e.g. YOLOv3~\cite{yolov3}, Mask R-CNN~\cite{maskrcnn}, etc.), and let $R = \{ (w_1, h_1), \ldots, (w_m, h_m) \}$ be the set of input resolutions. Then, during the caching phase, for each architecture $a_i$ and each input resolution $(w_j, h_j)$, we evaluate the execution time $t_{i,j}$ and validation accuracy $\alpha_{i,j}$ of the corresponding configuration, where parameters for other modules are taken from $\theta_\text{best}$. During tuning, given a configuration $\theta_k$ that uses architecture $a_i$ and input resolution $(w_j, h_j)$ for the detection module, we identify the choice of $a_{i'}$ and $(w_{j'}, h_{j'})$ with maximum $\alpha_{i',j'}$ such that $t_{i',j'} \le (1-S) t_{i,j}$, i.e., a new architecture and resolution that offers the highest accuracy among the choices that yield a speedup of at least $S$ over the previous selection.

\subsubsection{Proxy Model Module}

In the proxy model module, we need to pick two parameters: the model's input resolution, and the confidence threshold that specifies when the output score at a cell is high enough to mark that cell positive (and require object detector processing over the cell). One complication here is that speed is not directly related to the number of positive cells, since cells must first be grouped into rectangular windows where the object detector will be applied. On the other hand, if we can reduce the area of these windows by $S$, then this would speedup detector runtime by approximately $S$, and thus provide an overall speedup close to $S$.

We follow a similar approach as the detection module: during caching, we estimate the runtime and proxy model recall on the validation set at each resolution and threshold, and during tuning, we use these estimates to update parameters from the previous configuration. We define recall as the fraction of object detections that are covered by rectangular windows; during tuning, we will pick the resolution and threshold that have highest recall among choices with runtime at least $S$ faster than the previous configuration.

We first cache the per-cell proxy model classification scores for each resolution $(w_i, h_i)$ on every frame in the validation set, along with the detections computed by $\theta_\text{best}$. Let $T_{\text{proxy},i}$ be the runtime of the proxy model at each resolution. Then, for each threshold $B_j$, we compute rectangular windows using the cell grouping method from Section \ref{sec:msproxy} on each frame. Let $R_{i,j} = \{ r_1, \ldots, r_n \}$ be the set of rectangles computed across the frames at a given resolution and threshold. Then, our runtime estimate for this resolution and threshold is $T_{\text{proxy},i} + \sum_k T_{r_k.w, r_k.h}$, i.e., the proxy model runtime added to the detector runtime, which we estimate based on the rectangle sizes. The recall is the fraction of detections computed by $\theta_\text{best}$ that are covered by rectangles in $R_{i,j}$.

\subsubsection{Tracking Module}

The tracking module exposes a single parameter, the sampling gap $g$. We can obtain an overall $S$ speedup over the previous configuration by adjusting $g$ so that the tracker processes $S$ fewer frames. In particular, during tuning, we compute a new sampling gap by multiplying $g$ by $S$ and rounding up to the nearest power of two.

\section{Evaluation} \label{sec:eval}

We now evaluate MultiScope on 7 diverse video datasets against three baselines: Chameleon~\cite{chameleon}, BlazeIt~\cite{blazeit}, and Miris~\cite{miris}.
In our main results in Section \ref{sec:evalmain}, we evaluate the methods on the task of inferring all object tracks in video. While BlazeIt and Miris propose query-driven optimizations, here, we first apply them under their query-agnostic execution modes.
Then, in Section \ref{sec:evalblazeit}, we bring back these query-driven optimizations, and compare MultiScope and BlazeIt on processing cardinality-limited frame-level queries (a type of query that BlazeIt specifically optimizes for).

\smallskip
\noindent
\textbf{Datasets.} We conduct experiments on 7 diverse video datasets. We use the Tokyo, UAV, and Warsaw datasets from Miris~\cite{miris}, and the Amsterdam and Jackson datasets from BlazeIt~\cite{blazeit}. We also evaluate MultiScope on two new datasets, Caldot1 and Caldot2, consisting of video captured by California DOT cameras along two major highways. UAV consists of video captured from an aerial drone, while the other 6 datasets consist of video from fixed street-level cameras. Amsterdam captures activity at a riverside plaza, Caldot1 and Caldot2 capture highway activity, while the other 4 datasets capture city traffic junctions.

We sample three one-hour sets of 60 one-minute clips from each dataset: a training set (for training models), a validation set (for parameter tuning), and a hidden test set (for experimental evaluation). Thus, we allow each method to select Pareto-optimal configurations on the validation set, but report accuracy computed over a test set that the methods do not observe until execution.

\begin{figure}
\begin{center}
	\includegraphics[width=0.9\linewidth]{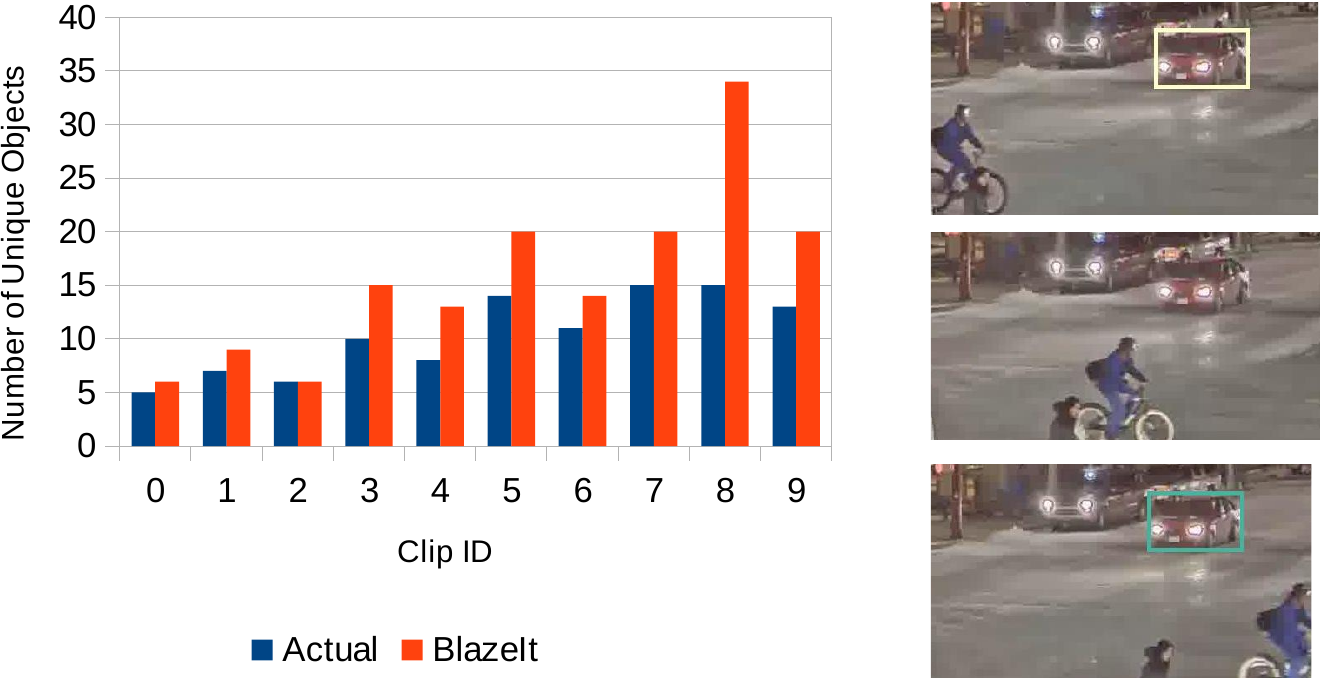}
\end{center}
	\caption{Comparison between BlazeIt's oracle configuration and actual hand-labeled ground truth on counting unique cars in 10 one-minute clips from the Jackson dataset. To the right, a sedan is split into two output tracks (manilla and blue) because it is not detected in intermediate frames.}
\label{fig:prelim1}
\end{figure}

\smallskip
\noindent
\textbf{Measuring Accuracy.}
In prior work (e.g., BlazeIt~\cite{blazeit}), the outputs of a best-accuracy configuration $\theta_\text{best}$ are used as a noisy oracle to evaluate approaches: the accuracy of a candidate configuration $\theta$ is measured by computing the relative similarity between its outputs and the outputs of $\theta_\text{best}$.
However, without hand-labeled ground truth, the degree of noise in the outputs of noisy oracles is unbounded. To demonstrate this, we hand-labeled the number of unique cars in ten uniformly sampled one-minute clips from the Jackson dataset, and compared these counts to the number of unique tracks produced by the BlazeIt oracle (which applies a detector and tracker at a high resolution and the native framerate). Here, we obtain BlazeIt tracks from CSV files published directly by the authors. We show results in Figure \ref{fig:prelim1}. The noisy oracle consistently over-estimates the number of cars, providing an average accuracy of only 72\% over the ten clips. Figure \ref{fig:prelim1} shows one source of errors, where one car is counted multiple times because it is missed by the detector in some frames. These results show that although the noisy oracle method reduces the amount of hand-labeling needed to execute a query, they hide an unbounded degree of error, and so accuracy measurements computed using relative comparisons to the noisy oracle are meaningless.

Instead, we hand-label ground truth data that consists of counts of unique objects (of a dataset-specific object category, which is cars in all 7 datasets) in each one-minute clip in the test set. In each clip, these counts are broken down by dataset-specific spatial patterns, where the number of patterns varies between 1 and 10 --- for example, in UAV, we annotate the number of cars that travel through a junction in each of eight observed turning directions (yielding 8 counts in each clip), while in Jackson, we simply count the number of unique cars (just 1 count per clip). Then, we measure accuracy by comparing the counts inferred from object tracks computed by a video analytics method with the ground truth counts, using percent accuracy averaged over both patterns and clips.

\begin{figure*}[t]
\begin{center}
	\includegraphics[width=\linewidth]{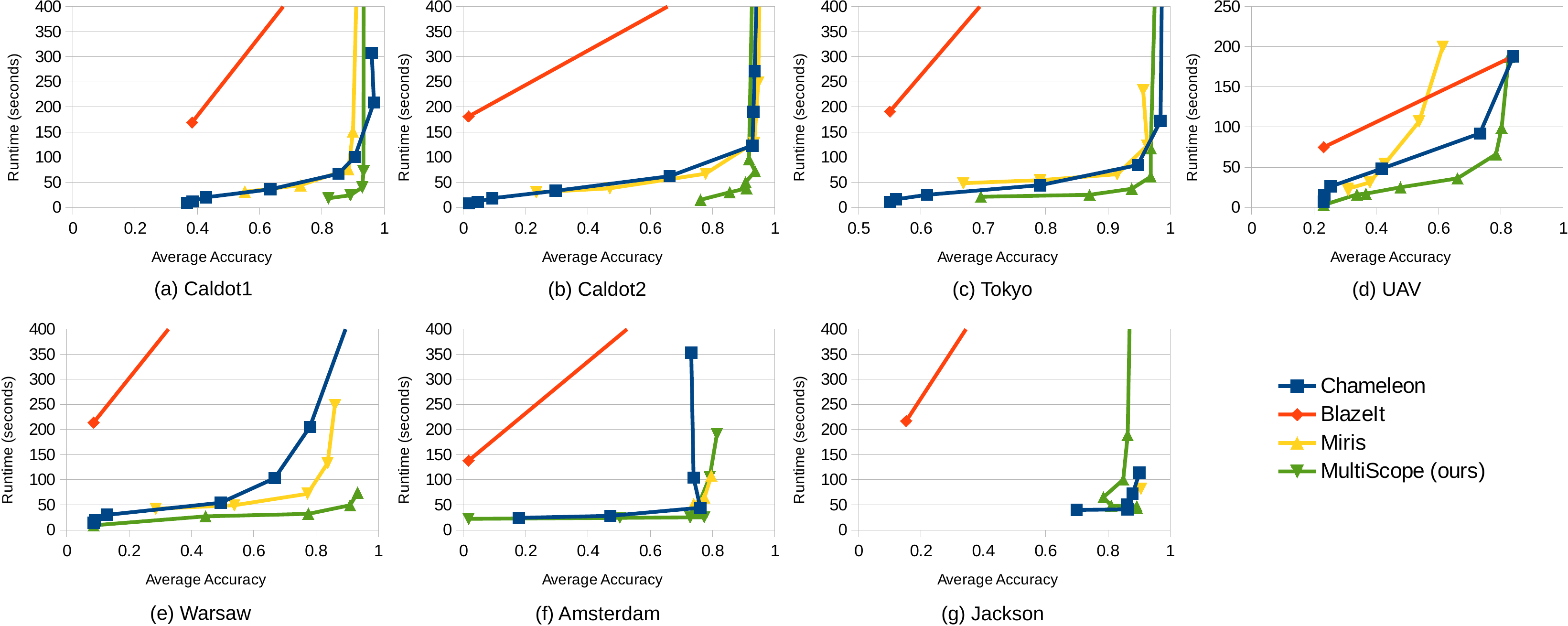}
\end{center}
	\caption{Runtime-accuracy curves on the test set of each dataset. Each point is a parameter configuration for the method, which is chosen based on performance in the validation set.}
\label{fig:results}
\end{figure*}

We also require an accuracy metric in the validation set for parameter tuning, and, in practice, for informing users about the performance of different candidate configurations. Like prior work, MultiScope could in principle tune parameters using a noisy oracle; however, as in the test set, this is impractical since the degree of noise in the resulting accuracy measurements would lead to poor and arbitrary decisions. Thus, we annotate counts in the one-hour validation set as well. This is a highly effective strategy for users to employ when applying MultiScope: annotation in the one-hour validation set requires only between 20 and 60 person-minutes for the 7 datasets we tested, and only needs to be conducted once per dataset.
Note that this is not a unique requirement of MultiScope: all baselines require accuracy measurements both for selecting a set of parameter configurations and for reporting the performance of each configuration to the user; noisy ground truth would yield poorly chosen configurations and would make it impossible for users to judge whether a proposed configuration will provide sufficient true accuracy. In fact, we argue that employing noisy oracles is a major flaw in the experimental evaluations of prior work. We extend all baselines (which we detail below) to use the count annotations during parameter tuning.

\smallskip
\noindent
\textbf{Baselines.} We compare MultiScope to three baselines. Chameleon~\cite{chameleon} explores performance under varying detector input resolutions and sampling rates to select the best parameters for a simple object detection pipeline.
BlazeIt~\cite{blazeit} applies classification and regression proxy models to optimize video analytics performance.
Miris~\cite{miris} applies variable framerate tracking under different error tolerances to provide tradeoffs between speed and accuracy.

In our main results (Section \ref{sec:evalmain}), we evaluate the methods on the task of extracting all tracks from video: while BlazeIt and Miris propose query-driven optimizations, they also have a query-agnostic execution mode, which we use here; in BlazeIt, this is a NoScope-like~\cite{noscope} mode where we skip detector processing on frames where the proxy model score falls below a confidence threshold (indicating that the frame likely contains no objects), and in Miris, this involves using a predicate that selects all tracks. Later, in Section \ref{sec:evalblazeit}, we will bring back these query-driven optimizations with an evaluation on processing cardinality-limited frame-level queries.

For consistency and simplicity, we use our implementations of the baselines. Each baseline includes a parameter selection phase where parameters are chosen that provide the best accuracy at several different throughput levels; we modify this phase in all baselines to utilize the count-based ground truth data and metric detailed above rather than ``ground truth'' computed automatically through a noisy oracle, since the latter is error-prone.

\smallskip
\noindent
\textbf{Implementation.} We store the training, validation, and test video on a local SSD at a fixed resolution, which is $720 \times 480$ for Caldot1 and Caldot2 and $1280 \times 720$ for the other 5 datasets, encoded in mp4 container with H264. The native video framerate ranges from 5 fps for UAV to 30 fps for Amsterdam and Jackson. The execution pipeline for all methods involves decoding the video using ffmpeg, and processing decoded frames iteratively in a way that is specific to each method. Frames are decoded at the object detector resolution, so reducing the model input resolution may provide a speedup both through faster model execution and through faster video decoding (ffmpeg incorporates optimizations to decode lower resolution frames more quickly, even though the source resolution is fixed). After the last frame is decoded and processed, each method outputs a set of extracted object tracks. The parameter selection phase for all methods involves repeatedly applying the execution pipeline with different parameter configurations.

\subsection{Results} \label{sec:evalmain}

We now evaluate MultiScope against the 3 baselines on each of the 7 datasets. We measure runtime on a machine with one NVIDIA Tesla V100 GPU and one Intel Xeon Gold 6142 CPU, and exclude training costs of each method from the runtime, focusing only on costs that grow linearly with the dataset size. We first apply the parameter selection phase of each method on the validation set, which yields a speed-accuracy curve consisting of several Pareto-optimal configurations. We then apply each of these parameter configurations on the unseen test set, yielding a test speed-accuracy curve: the runtime of a configuration is its execution time over the one-hour test set, and its accuracy is the percentage accuracy of its output counts in each one-minute clip, averaged across 60 clips.

Figure \ref{fig:results} shows the results. To simplify the comparison between the methods, we also summarize the results in Table \ref{tab:results}: for each dataset and method, we show the runtime of the fastest configuration (among candidates selected using the validation set) providing accuracy within 5\% of the best achieved accuracy on the test set (across all methods). We choose the 5\% threshold since average accuracy is computed over 60 accuracy samples (the test set consists of 60 clips randomly sampled from the video dataset), and so the variance in the sample mean implies that differences in accuracy less than 5\% may not be meaningful. Thus, Table \ref{tab:results} highlights the runtime that each method can achieve while maintaining an accuracy close to the best accuracy achieved by any method.

\begin{table}[t]
    \centering
    \caption{Runtime in seconds of each method on the test set of each dataset, using the fastest candidate configuration that provides accuracy within 5\% of the best achieved accuracy.
    }
    \begin{tabular}{|l|r|r|r|r|}
        \hline
        \textbf{Dataset} & \textbf{MultiScope} & \textbf{Chameleon} & \textbf{BlazeIt} & \textbf{Miris} \\
        \hline
        Caldot1 & \textbf{40} & 209 & 990 & 533 \\
        \hline
        Caldot2 & \textbf{38} & 123 & 803 & 129 \\
        \hline
        Tokyo & \textbf{37} & 84 & 823 & 123 \\
        \hline
        UAV & \textbf{99} & 188 & 323 & - \\
        \hline
        Warsaw & \textbf{49} & 422 & 867 & 249 \\
        \hline
        Amsterdam & \textbf{25} & 44 & 666 & 64 \\
        \hline
        Jackson & 44 & \textbf{41} & 618 & 82 \\
        \hline
    \end{tabular}
    \label{tab:results}
\end{table}

MultiScope consistently performs comparably to or better than the next best baseline across all seven datasets: it provides an average 2.9x speedup over the next best baseline in Table \ref{tab:results}. Although Chameleon and Miris offer good speed-accuracy tradeoffs on every dataset, MultiScope provides better performance (especially on Tokyo, UAV, Warsaw, Caldot1, and Caldot2), both through the novel techniques employed in its segmentation proxy model and recurrent tracker components, and simply by exploring a more diverse range of ways to improve execution speed without impairing accuracy. BlazeIt provides some degree of a tradeoff between speed and accuracy on Amsterdam, Caldot1, and Caldot2; however, it only yields two candidate configurations for the other datasets because those datasets have objects visible in every frame (one candidate applies the detector on every frame, while the other skips the entire video and simply outputs 0 for all counts).
Across all accuracy levels, MultiScope is better than all other methods on 5 out of 7 datasets; on Amsterdam and Jackson, it provides a comparable speed-accuracy curve to Chameleon and Miris.

\smallskip
\noindent
\textbf{Other Costs.}
Here, we have focused on runtime costs that scale linearly with the dataset size, ignoring sub-linear parameter tuning and model training costs --- e.g., training time is constant for a particular model architecture and dataset, while tuning scales with the square root of the dataset size (to maintain a fixed sampling error bound). On Caldot1, MultiScope uses 19 minutes to sample the one-hour training and validation sets, 4 minutes to train the segmentation proxy models, 8 minutes to train the tracker model, and 48 minutes for parameter tuning.

\begin{figure}
\begin{center}
	\includegraphics[width=0.8\linewidth]{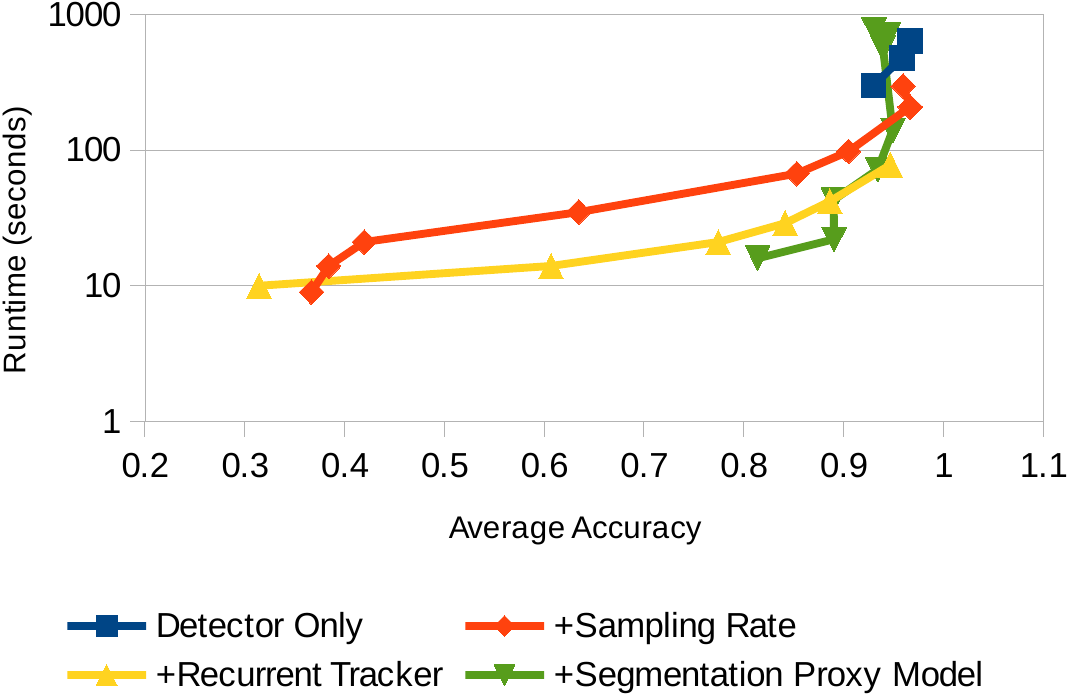}
\end{center}
	\caption{Ablation study of MultiScope.}
\label{fig:ablation}
\end{figure}

\smallskip
\noindent
\textbf{Ablation Study.}
In Figure \ref{fig:ablation}, we conduct an ablation study of MultiScope on the Caldot1 test set, testing four successively more complete implementations of MultiScope: in blue we start with our parameter tuning method with only the object detection module;
in red we add a tracking module using the heuristic SORT~\cite{sort} tracker; in yellow we replace SORT with our recurrent reduced-rate tracking method; and in green we add the segmentation proxy model, which corresponds to our full method. Each additional module improves performance along a portion of the resulting speed-accuracy curve.

\subsection{MultiScope over Query-Driven Methods} \label{sec:evalblazeit}

\begin{table}[t]
    \centering
    \caption{Comparison between BlazeIt and MultiScope on finding 20 frames in the Jackson dataset with at least 4 cars.}
    \begin{tabular}{|l|r|r|}
        \hline
        Method & BlazeIt & MultiScope \\
        \hline
        Pre-processing Time (sec) & 165 & 184 \\
        \hline
        Query Time (sec) & 44 & 1 \\
        \hline
        Total Time (sec) & 209 & 185 \\
        \hline
        Accuracy & 18/20 & 20/20 \\
        \hline
    \end{tabular}
    \label{tab:blazeit}
\end{table}

In the previous section, we applied BlazeIt and Miris in their query-agnostic execution modes. However, query-time optimizations are a major component of these methods.
We now show that MultiScope performs comparably to the query-driven method in BlazeIt for optimizing limit query execution, even in a scenario where BlazeIt is executed to extract just 20 matching frames, while MultiScope extracts all tracks from the dataset (after which the query can be efficiently processed over the computed tracks).
We apply MultiScope and BlazeIt to extract any 20 frames from a five-hour subset of the Jackson dataset where there are at least four cars in the bottom half of the video. We also require the output frames to be spaced at least five seconds apart from one another. We focus on the bottom half of the video to avoid ambiguity issues associated with the horizon. As before, we ignore training costs in both methods, and focus only on execution runtime that grows linearly with the dataset size.

To apply BlazeIt, we train the proxy model to count the number of cars in the bottom half of the input frame, and then compute the count estimates on every frame in the five-hour dataset. Since the proxy model operates on $64 \times 64$ frames, the cost of this phase is dominated by video decoding (which is fast since we use ffmpeg to directly decode $64 \times 64$ frames). BlazeIt optimizes this limit query by applying the detector (using an input resolution that provides the best accuracy, in our implementation) on frames in order from highest-scoring to lowest-scoring until it encounters 20 frames where the detector finds at least four cars in the bottom half.

To apply MultiScope, we first simply execute MultiScope to extract all tracks from the dataset, using the fastest parameter configuration among the candidates that yield accuracy within 5\% of the best-achieved accuracy. We then post-process the tracks by picking 20 frames with at least four detections in the bottom half, starting with frames where the tracks visible in the bottom half have the highest minimum duration.

Table \ref{tab:blazeit} shows the results. The pre-processing time in both methods (i.e., proxy model runtime in BlazeIt and track extraction time in MultiScope) is dominated by video decoding. Thus, despite its extensive limit query optimizations, BlazeIt provides no speedup over MultiScope on this query, since it must first apply the proxy model over the entire dataset. BlazeIt's query runtime corresponds to executing the object detector on 3,674 frames before finding 20 outputs; we exclude video decoding time in this number since we have not optimized the query pipeline for random-access decoding. We measure accuracy by manually confirming how many of the 20 frames output by each method actually contain four cars in the bottom half. Two frames produced by BlazeIt contain spurious detections, which MultiScope avoids by ignoring tracks that consist of only a single detection. Nevertheless, the main conclusion from this experiment is that MultiScope is able to accurately extract all tracks from video in the same time that prior work requires for answering a single limit query, even when the desired output cardinality is small. After tracks are inferred, exploratory queries can be executed in milliseconds with MultiScope instead of minutes with BlazeIt.

\subsection{Count Accuracy vs MOTA} \label{sec:evalmota}

\begin{figure}
\begin{center}
	\includegraphics[width=\linewidth]{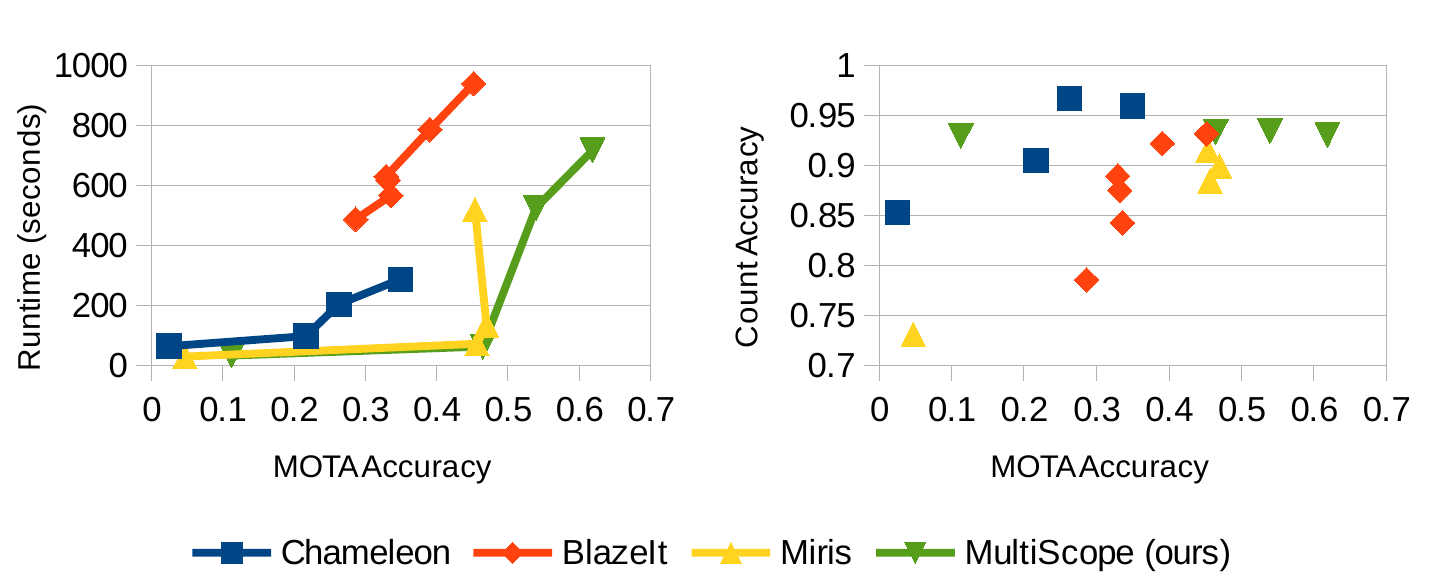}
\end{center}
	\caption{On the left, MOTA accuracy and runtime curves on the Caldot1 test set. On the right, a scatterplot showing the correlation between Count Accuracy and MOTA over candidate parameter configurations across methods.}
\label{fig:mota}
\end{figure}

Up to now, we have focused on a count-based accuracy metric over a hidden test set that corresponds to the user-provided labels on the validation set. It is possible, though, that a method that scores highly in terms of count accuracy does not actually provide high-quality tracks. In this section, we show that this is not the case, and that count accuracy and track quality are highly correlated.

To do so, we label bounding boxes in every frame of 4 one-minute clips in the Caldot1 test set at 1 fps, along with track IDs for each box. We then compare tracks inferred by MultiScope and the baselines against the ground truth labels in terms of Multi-Object Tracking Accuracy (MOTA)~\cite{mot16}, which is a standard metric from the computer vision community for comprehensively evaluating the accuracy of inferred tracks. MOTA considers several factors, including whether inferred bounding boxes match with ground truth boxes, and whether ground truth tracks are split into multiple inferred tracks or vice versa.

In Figure \ref{fig:mota}, we compute the MOTA of each candidate configuration produced by each method over the 4 labeled Caldot1 clips. The candidate configurations are derived by each method by optimizing for count accuracy on the validation set.
On the left, we plot MOTA against runtime, and on the right, we plot MOTA against the corresponding count accuracy. The right chart shows that MOTA is highly correlated with count accuracy across all of the methods, while the left chart validates that MultiScope is effective at not only recovering counts, but also at robustly extracting object tracks.

\section{Related Work}

Several systems have recently been proposed for performing video analytics tasks over large-scale video datasets. NoScope~\cite{noscope} and Probabilistic Predicates~\cite{probpred} propose training a classification proxy model to input low-resolution video frames and quickly determine whether the frame is relevant to the query (e.g., whether the frame contains any detections of a particular object category); processing of expensive machine learning models such as object detectors can be skipped on frames where the proxy model is confident it does not match the query. Video Monitoring Queries~\cite{vmq}, Focus~\cite{focus}, and several other works propose various extensions to the proxy model technique. In particular, BlazeIt~\cite{blazeit} employs an improved proxy model architecture that provides higher accuracy, and proposes techniques specialized for efficiently executing limit and aggregate queries by post-processing the proxy model outputs. Other techniques for optimizing video analytics tasks have been explored as well. Chameleon~\cite{chameleon} proposes optimizing the object detector input resolution and sampling framerate to robustly extract detections from a video dataset. Miris~\cite{miris} proposes a variable rate tracking method that processes video at substantially reduced framerates when possible to accurately answer object track queries.

Our work is also related to two techniques explored extensively in the computer vision community: multi-scale detection and recurrent multi-object tracking. MultiScope's core contribution is improving video query optimizations proposed in the data management community by adapting and integrating these computer vision techniques. Dynamic Zoom-in Networks~\cite{multiscale1} and AutoFocus~\cite{multiscale2} propose to optimize object detection speed with a coarse-to-fine architecture, where objects are first detected in a down-sampled image, and portions of an image are processed at higher resolutions if another model predicts this will yield a high accuracy gain. Recent recurrent multi-object tracking methods include Deep Tracklet Association~\cite{rnnmot1} and Bilinear-LSTM~\cite{rnnmot2}.

\section{Conclusion}

In this paper, we have presented MultiScope, a video pre-processor for exploratory video analytics queries. Compared to prior work, MultiScope is faster, offering a 2.9x average speedup across 7 datasets over the next best baseline at the same accuracy level; is more general, enabling execution of any query that involves object detections and tracks; and substantially reduces per-query execution time after pre-processing, since queries can be answered by processing extracted tracks without additional video decoding and ML inference.

\bibliographystyle{ACM-Reference-Format}
\bibliography{main}


\begin{thebibliography}{20}


\ifx \showCODEN    \undefined \def \showCODEN     #1{\unskip}     \fi
\ifx \showDOI      \undefined \def \showDOI       #1{#1}\fi
\ifx \showISBNx    \undefined \def \showISBNx     #1{\unskip}     \fi
\ifx \showISBNxiii \undefined \def \showISBNxiii  #1{\unskip}     \fi
\ifx \showISSN     \undefined \def \showISSN      #1{\unskip}     \fi
\ifx \showLCCN     \undefined \def \showLCCN      #1{\unskip}     \fi
\ifx \shownote     \undefined \def \shownote      #1{#1}          \fi
\ifx \showarticletitle \undefined \def \showarticletitle #1{#1}   \fi
\ifx \showURL      \undefined \def \showURL       {\relax}        \fi
\providecommand\bibfield[2]{#2}
\providecommand\bibinfo[2]{#2}
\providecommand\natexlab[1]{#1}
\providecommand\showeprint[2][]{arXiv:#2}

\bibitem[\protect\citeauthoryear{Bastani, He, Balasingam, Gopalakrishnan,
  Alizadeh, Balakrishnan, Cafarella, Kraska, and Madden}{Bastani
  et~al\mbox{.}}{2020}]%
        {miris}
\bibfield{author}{\bibinfo{person}{Favyen Bastani}, \bibinfo{person}{Songtao
  He}, \bibinfo{person}{Arjun Balasingam}, \bibinfo{person}{Karthik
  Gopalakrishnan}, \bibinfo{person}{Mohammad Alizadeh}, \bibinfo{person}{Hari
  Balakrishnan}, \bibinfo{person}{Michael Cafarella}, \bibinfo{person}{Tim
  Kraska}, {and} \bibinfo{person}{Sam Madden}.}
  \bibinfo{year}{2020}\natexlab{}.
\newblock \showarticletitle{{MIRIS: Fast Object Track Queries in Video}}. In
  \bibinfo{booktitle}{\emph{Proceedings of the 2020 ACM SIGMOD International
  Conference on Management of Data}}. \bibinfo{pages}{1907--1921}.
\newblock


\bibitem[\protect\citeauthoryear{Bewley, Ge, Ott, Ramos, and Upcroft}{Bewley
  et~al\mbox{.}}{2016}]%
        {sort}
\bibfield{author}{\bibinfo{person}{Alex Bewley}, \bibinfo{person}{Zongyuan Ge},
  \bibinfo{person}{Lionel Ott}, \bibinfo{person}{Fabio Ramos}, {and}
  \bibinfo{person}{Ben Upcroft}.} \bibinfo{year}{2016}\natexlab{}.
\newblock \showarticletitle{{Simple Online and Realtime Tracking}}. In
  \bibinfo{booktitle}{\emph{IEEE International Conference on Image Processing
  (ICIP)}}. IEEE, \bibinfo{pages}{3464--3468}.
\newblock


\bibitem[\protect\citeauthoryear{Chin, Ding, and Marculescu}{Chin
  et~al\mbox{.}}{2019}]%
        {adascale}
\bibfield{author}{\bibinfo{person}{Ting-Wu Chin}, \bibinfo{person}{Ruizhou
  Ding}, {and} \bibinfo{person}{Diana Marculescu}.}
  \bibinfo{year}{2019}\natexlab{}.
\newblock \showarticletitle{{AdaScale: Towards Real-time Video Object Detection
  Using Adaptive Scaling}}. In \bibinfo{booktitle}{\emph{Systems and Machine
  Learning Conference (SysML)}}.
\newblock


\bibitem[\protect\citeauthoryear{Chu and Ling}{Chu and Ling}{2019}]%
        {famnet}
\bibfield{author}{\bibinfo{person}{Peng Chu} {and} \bibinfo{person}{Haibin
  Ling}.} \bibinfo{year}{2019}\natexlab{}.
\newblock \showarticletitle{{FAMNet: Joint Learning of Feature, Affinity and
  Multi-dimensional Assignment for Online Multiple Object Tracking}}. In
  \bibinfo{booktitle}{\emph{IEEE International Conference on Computer Vision
  (ICCV)}}.
\newblock


\bibitem[\protect\citeauthoryear{Gao, Yu, Li, Morariu, and Davis}{Gao
  et~al\mbox{.}}{2018}]%
        {multiscale1}
\bibfield{author}{\bibinfo{person}{Mingfei Gao}, \bibinfo{person}{Ruichi Yu},
  \bibinfo{person}{Ang Li}, \bibinfo{person}{Vlad~I Morariu}, {and}
  \bibinfo{person}{Larry~S Davis}.} \bibinfo{year}{2018}\natexlab{}.
\newblock \showarticletitle{Dynamic zoom-in network for fast object detection
  in large images}. In \bibinfo{booktitle}{\emph{Proceedings of the IEEE
  Conference on Computer Vision and Pattern Recognition}}.
  \bibinfo{pages}{6926--6935}.
\newblock


\bibitem[\protect\citeauthoryear{He, Gkioxari, Doll{\'a}r, and Girshick}{He
  et~al\mbox{.}}{2017}]%
        {maskrcnn}
\bibfield{author}{\bibinfo{person}{Kaiming He}, \bibinfo{person}{Georgia
  Gkioxari}, \bibinfo{person}{Piotr Doll{\'a}r}, {and} \bibinfo{person}{Ross
  Girshick}.} \bibinfo{year}{2017}\natexlab{}.
\newblock \showarticletitle{{Mask R-CNN}}. In \bibinfo{booktitle}{\emph{IEEE
  International Conference on Computer Vision (ICCV)}}.
  \bibinfo{pages}{2961--2969}.
\newblock


\bibitem[\protect\citeauthoryear{Hsieh, Ananthanarayanan, Bodik, Venkataraman,
  Bahl, Philipose, Gibbons, and Mutlu}{Hsieh et~al\mbox{.}}{2018}]%
        {focus}
\bibfield{author}{\bibinfo{person}{Kevin Hsieh}, \bibinfo{person}{Ganesh
  Ananthanarayanan}, \bibinfo{person}{Peter Bodik}, \bibinfo{person}{Shivaram
  Venkataraman}, \bibinfo{person}{Paramvir Bahl}, \bibinfo{person}{Matthai
  Philipose}, \bibinfo{person}{Phillip~B Gibbons}, {and} \bibinfo{person}{Onur
  Mutlu}.} \bibinfo{year}{2018}\natexlab{}.
\newblock \showarticletitle{{Focus: Querying large video datasets with low
  latency and low cost}}. In \bibinfo{booktitle}{\emph{13th $\{$USENIX$\}$
  Symposium on Operating Systems Design and Implementation ($\{$OSDI$\}$ 18)}}.
  \bibinfo{pages}{269--286}.
\newblock


\bibitem[\protect\citeauthoryear{Jiang, Ananthanarayanan, Bodik, Sen, and
  Stoica}{Jiang et~al\mbox{.}}{2018}]%
        {chameleon}
\bibfield{author}{\bibinfo{person}{Junchen Jiang}, \bibinfo{person}{Ganesh
  Ananthanarayanan}, \bibinfo{person}{Peter Bodik}, \bibinfo{person}{Siddhartha
  Sen}, {and} \bibinfo{person}{Ion Stoica}.} \bibinfo{year}{2018}\natexlab{}.
\newblock \showarticletitle{{Chameleon: Scalable Adaptation of Video
  Analytics}}. In \bibinfo{booktitle}{\emph{Proceedings of the 2018 Conference
  of the ACM Special Interest Group on Data Communication}}.
  \bibinfo{pages}{253--266}.
\newblock


\bibitem[\protect\citeauthoryear{Kang, Bailis, and Zaharia}{Kang
  et~al\mbox{.}}{2019}]%
        {blazeit}
\bibfield{author}{\bibinfo{person}{Daniel Kang}, \bibinfo{person}{Peter
  Bailis}, {and} \bibinfo{person}{Matei Zaharia}.}
  \bibinfo{year}{2019}\natexlab{}.
\newblock \showarticletitle{{Challenges and Opportunities in {DNN}-Based Video
  Analytics: A Demonstration of the BlazeIt Video Query Engine}}. In
  \bibinfo{booktitle}{\emph{Conference on Innovative Data Systems Research
  (CIDR)}}.
\newblock


\bibitem[\protect\citeauthoryear{Kang, Emmons, Abuzaid, Bailis, and
  Zaharia}{Kang et~al\mbox{.}}{2017}]%
        {noscope}
\bibfield{author}{\bibinfo{person}{Daniel Kang}, \bibinfo{person}{John Emmons},
  \bibinfo{person}{Firas Abuzaid}, \bibinfo{person}{Peter Bailis}, {and}
  \bibinfo{person}{Matei Zaharia}.} \bibinfo{year}{2017}\natexlab{}.
\newblock \showarticletitle{{NoScope: Optimizing Neural Network Queries over
  Video at Scale}}. In \bibinfo{booktitle}{\emph{Proceedings of the VLDB
  Endowment}}.
\newblock


\bibitem[\protect\citeauthoryear{Kim, Li, and Rehg}{Kim et~al\mbox{.}}{2018}]%
        {rnnmot2}
\bibfield{author}{\bibinfo{person}{Chanho Kim}, \bibinfo{person}{Fuxin Li},
  {and} \bibinfo{person}{James~M Rehg}.} \bibinfo{year}{2018}\natexlab{}.
\newblock \showarticletitle{{Multi-Object Tracking with Neural Gating using
  Bilinear LSTM}}. In \bibinfo{booktitle}{\emph{Proceedings of the European
  Conference on Computer Vision (ECCV)}}. \bibinfo{pages}{200--215}.
\newblock


\bibitem[\protect\citeauthoryear{Koudas, Li, and Xarchakos}{Koudas
  et~al\mbox{.}}{2020}]%
        {vmq}
\bibfield{author}{\bibinfo{person}{Nick Koudas}, \bibinfo{person}{Raymond Li},
  {and} \bibinfo{person}{Ioannis Xarchakos}.} \bibinfo{year}{2020}\natexlab{}.
\newblock \showarticletitle{{Video Monitoring Queries}}. In
  \bibinfo{booktitle}{\emph{2020 IEEE 36th International Conference on Data
  Engineering (ICDE)}}. IEEE, \bibinfo{pages}{1285--1296}.
\newblock


\bibitem[\protect\citeauthoryear{Liu, Chu, Liu, and Yu}{Liu
  et~al\mbox{.}}{2020}]%
        {gsm}
\bibfield{author}{\bibinfo{person}{Qiankun Liu}, \bibinfo{person}{Qi Chu},
  \bibinfo{person}{Bin Liu}, {and} \bibinfo{person}{Nenghai Yu}.}
  \bibinfo{year}{2020}\natexlab{}.
\newblock \showarticletitle{{GSM: Graph Similarity Model for Multi-Object
  Tracking}}. In \bibinfo{booktitle}{\emph{International Joint Conference on
  Artificial Intelligence (IJCAI)}}.
\newblock


\bibitem[\protect\citeauthoryear{Lu, Chowdhery, Kandula, and Chaudhuri}{Lu
  et~al\mbox{.}}{2018}]%
        {probpred}
\bibfield{author}{\bibinfo{person}{Yao Lu}, \bibinfo{person}{Aakanksha
  Chowdhery}, \bibinfo{person}{Srikanth Kandula}, {and}
  \bibinfo{person}{Surajit Chaudhuri}.} \bibinfo{year}{2018}\natexlab{}.
\newblock \showarticletitle{{Accelerating Machine Learning Inference with
  Probabilistic Predicates}}. In \bibinfo{booktitle}{\emph{International
  Conference on Management of Data (SIGMOD)}}. ACM,
  \bibinfo{pages}{1493--1508}.
\newblock


\bibitem[\protect\citeauthoryear{Milan, Leal-Taix{\'e}, Reid, Roth, and
  Schindler}{Milan et~al\mbox{.}}{2016}]%
        {mot16}
\bibfield{author}{\bibinfo{person}{Anton Milan}, \bibinfo{person}{Laura
  Leal-Taix{\'e}}, \bibinfo{person}{Ian Reid}, \bibinfo{person}{Stefan Roth},
  {and} \bibinfo{person}{Konrad Schindler}.} \bibinfo{year}{2016}\natexlab{}.
\newblock \showarticletitle{{MOT16: A Benchmark for Multi-Object Tracking}}.
\newblock \bibinfo{journal}{\emph{arXiv preprint arXiv:1603.00831}}
  (\bibinfo{year}{2016}).
\newblock


\bibitem[\protect\citeauthoryear{Najibi, Singh, and Davis}{Najibi
  et~al\mbox{.}}{2019}]%
        {multiscale2}
\bibfield{author}{\bibinfo{person}{Mahyar Najibi}, \bibinfo{person}{Bharat
  Singh}, {and} \bibinfo{person}{Larry~S Davis}.}
  \bibinfo{year}{2019}\natexlab{}.
\newblock \showarticletitle{{AutoFocus: Efficient Multi-scale Inference}}. In
  \bibinfo{booktitle}{\emph{Proceedings of the IEEE International Conference on
  Computer Vision}}. \bibinfo{pages}{9745--9755}.
\newblock


\bibitem[\protect\citeauthoryear{Redmon, Divvala, Girshick, and Farhadi}{Redmon
  et~al\mbox{.}}{2016}]%
        {yolo}
\bibfield{author}{\bibinfo{person}{Joseph Redmon}, \bibinfo{person}{Santosh
  Divvala}, \bibinfo{person}{Ross Girshick}, {and} \bibinfo{person}{Ali
  Farhadi}.} \bibinfo{year}{2016}\natexlab{}.
\newblock \showarticletitle{{You Only Look Once: Unified, Real-Time Object
  Detection}}. In \bibinfo{booktitle}{\emph{IEEE Conference on Computer Vision
  and Pattern Recognition (CVPR)}}.
\newblock


\bibitem[\protect\citeauthoryear{Redmon and Farhadi}{Redmon and
  Farhadi}{2018}]%
        {yolov3}
\bibfield{author}{\bibinfo{person}{Joseph Redmon} {and} \bibinfo{person}{Ali
  Farhadi}.} \bibinfo{year}{2018}\natexlab{}.
\newblock \bibinfo{booktitle}{\emph{{YOLOv3: An Incremental Improvement}}}.
\newblock \bibinfo{type}{{T}echnical {R}eport}.
  \bibinfo{institution}{University of Washington}.
\newblock


\bibitem[\protect\citeauthoryear{Zhang, Sheng, Wu, Wang, Lyu, Ke, and
  Xiong}{Zhang et~al\mbox{.}}{2020}]%
        {rnnmot1}
\bibfield{author}{\bibinfo{person}{Yang Zhang}, \bibinfo{person}{Hao Sheng},
  \bibinfo{person}{Yubin Wu}, \bibinfo{person}{Shuai Wang},
  \bibinfo{person}{Weifeng Lyu}, \bibinfo{person}{Wei Ke}, {and}
  \bibinfo{person}{Zhang Xiong}.} \bibinfo{year}{2020}\natexlab{}.
\newblock \showarticletitle{{Long-term Tracking with Deep Tracklet
  Association}}.
\newblock \bibinfo{journal}{\emph{IEEE Transactions on Image Processing}}
  (\bibinfo{year}{2020}).
\newblock


\bibitem[\protect\citeauthoryear{Zhou, Koltun, and Kr{\"a}henb{\"u}hl}{Zhou
  et~al\mbox{.}}{2020}]%
        {cttrack}
\bibfield{author}{\bibinfo{person}{Xingyi Zhou}, \bibinfo{person}{Vladlen
  Koltun}, {and} \bibinfo{person}{Philipp Kr{\"a}henb{\"u}hl}.}
  \bibinfo{year}{2020}\natexlab{}.
\newblock \showarticletitle{{Tracking Objects as Points}}. In
  \bibinfo{booktitle}{\emph{European Conference on Computer Vision (ECCV)}}.
\newblock


\end{thebibliography}

\end{document}